\documentclass[floatfix, twocolumn,amsmath,amssymb, aps,prb,superscriptaddress]{revtex4-2}
\usepackage{mathrsfs}
\usepackage{amsmath}
\usepackage{ragged2e}

\usepackage{caption}
\usepackage{subcaption}
\usepackage{graphicx}
\usepackage{hyperref}
\usepackage{stfloats}
\usepackage{ulem}
\hypersetup{ colorlinks, citecolor=blue, linkcolor=blue, urlcolor=blue}

\DeclareCaptionLabelFormat{nocaption}{}
\graphicspath{{img/}}

\begin{document}
	\title{Topological features of Haldane model on a dice lattice: role of flat band on transport properties}
	
	\author{Sayan Mondal}
	\affiliation{Department of Physics, Indian Institute of Technology Guwahati, 
		Guwahati 781039, Assam, India}
	\author{Saurabh Basu}
	\affiliation{Department of Physics, Indian Institute of Technology Guwahati, 
		Guwahati 781039, Assam, India}

	\begin{abstract}
	
	We study the topological properties of a Haldane model on a band deformed dice lattice, which has three atoms per unit cell (call them as A, B and C) and the spectrum comprises of three bands, including a flat band. The bands are systematically deformed with an aim to study the evolution of topology and the transport properties. The deformations are induced through hopping anisotropies and are achieved in two distinct ways. In one of them, the hopping amplitudes between the sites of B and C sublattices and those between A and B sublattices are varied along a particular direction, and in the other, the hopping between the sites of A and B sublattices are varied (keeping B-C hopping unaltered) along the same direction. The first case retains some of the spectral features of the familiar dice lattice and yields Chern insulating lobes in the phase diagram with $C=\pm2$ till a certain critical deformation. The topological features are supported by the presence of a pair of chiral edge modes at each edge of a ribbon and the plateaus observed in the anomalous Hall conductivity support the above scenario. Whereas, a selective tuning of only the A-B hopping amplitudes distorts the flat band and has important ramifications on the topological properties of the system. The insulating lobes in the phase diagram have distinct features compared to the case above, and there are dips observed in the Hall conductivity near the zero bias. The dip widens as the hopping anisotropy is made larger, and thus the scenario registers significant deviation from the familiar plateau structure observed in the anomalous Hall conductivity. However, a phase transition from a topological to a trivial insulating region demonstrated by the Chern number changing discontinuously from $\pm2$ to zero beyond a certain critical hopping anisotropy remains a common feature in the two cases.

	\end{abstract}
	\maketitle
	\section{Introduction}\label{sec:intro}
	Since the discovery of quantum Hall effect (QHE) in 1980 \cite{klitzing1980}, there has been a surge of research on the topological phases of matter in the condensed matter community. The Hall conductivity in the QHE shows a series of plateaus \cite{laughlin, trugman1983, ilani2004, vasil1985} quantized in units of $e^2/h$ and the topological invariant which describes this quantization is known as the Thouless–Kohmoto–Nightingale–den Nijs (TKNN) invariant \cite{thouless1982}. Apart from the Hall effect observed in presence of a magnetic field, there has been proposal to realize such a behaviour in the absence of magnetic field, namely, the quantum anomalous Hall effect (QAHE) \cite{chang2013, chang2015, deng2020, nagaosa2010} that relies on the breaking of time-reversal symmetry (TRS) of the system. Haldane pioneered such an idea in a two-dimensional honeycomb lattice \cite{haldane1988}. To break the TRS, he proposed a complex next-nearest-neighbour (NNN) hopping with a phase $\phi$. In this model, the bands are associated with a topological invariant, called the Chern number, which is identical with the quantization of the Hall conductivity. Further, a sublattice symmetry breaking term, which we shall call later as the Semenoff mass (denoted by $\Delta$) is included, which induces opening or closing of the spectral gap in the band structure. The variation of the Chern number as a function, $\Delta$ and $\phi$ presents a phase diagram which encodes these opening and closing of band gaps at the Dirac points (usually denoted by the $\mathbf{K}$ and $\mathbf{K^\prime}$ points) in the Brillouin zone (BZ). Further, the value of the Chern number ascertains the nature of the gap, for example, if it is finite (zero), it denotes a topological (trivial) gap.
	
	Following Haldane's work in 1988, there has been of immense interest on the QAHE in several two-dimensional systems, such as, the Lieb lattice \cite{weeks2010, apaja2010, goldman2011, tsai2015}, checkerboard lattice \cite{sun2009}, Kagom\'e lattice \cite{ohgushi2000, xiao2003, guo2009, liu2013} etc.  Experimentally a Haldane model has been realized in two dimensional honeycomb structures, such as, Fe-based ferromagnetic insulators, $X\mathrm{Fe_2(PO_4)_2}$, where $X$ may be K, Cs, La \cite{kim2017}. Further, cold atoms in an optical lattice created by standing-wave laser beams \cite{shao2008, alba2011, tarruel2012}, for example, an optical honeycomb lattice \cite{jotzu2014} also depicts similar non-trivial topological phases.
	
	Most of these concepts are applied to two band systems, where the Chern number assumes values $\pm1$. Hence, to serve a dual purpose of extending the calculations to other systems with larger number of bands (specifically a flat band), along with achieving higher values of the Chern number for a system, we consider a dice lattice. A dice lattice has the structure of a honeycomb lattice with an additional lattice point at the centre of the hexagon, which is connected to either of the A or B sublattice. This additional lattice point belongs to a third sublattice C \cite{sutherland1986, vidal1998, korsunov2001, rizzi2006, bercioux2009, vigh2013, malcolm2016, demler2006,urban2011,cooper2012}. Thus, the unit cell contains three sublattices, namely, A, B and C. The band structure consists of a zero energy flat band that resides between the upper (conduction) and the lower (valence) bands, and they have degeneracies at the Dirac points ($\mathbf{K}$ and $\mathbf{K^\prime}$) in the sense that all the bands touch each other at these points in the BZ.
	
	There is in general a great deal of excitement with systems possessing flat bands \cite{sutherland1986,aoki1996,khomeriki2016}. A flat band is dissipationless and provides an ideal platform for exploring electronic correlations owing to a complete quenching of kinetic energy of the particles. In a way, they are analogous to the Landau levels. There have been theoretical predictions about several intriguing phenomena, such as, Wigner crystallization in honeycomb structures \cite{wu2007}, large superconducting critical temperature \cite{volovik2016}, fractional Chern insulators \cite{tang2011,sun2011} etc. On the experimental front, flat bands have been achieved in photonic crystals \cite{mukherjee2015}, optical lattices \cite{hyrkhas2013} and metamaterials \cite{nakata2012}. Interestingly, in twisted bilayer graphene at the magic angle, the excitement of a close resemblance with the phase diagram of high-$T_c$ cuprates is due to the presence of almost  flat bands in the spectrum \cite{cao2018}.
	
	In this work, we wish to explore topological properties of flat band systems.	 
	 Gapped flat band systems demonstrate higher values of Chern number and they imply larger values of the (anomalous) Hall conductivity \cite{kapri2020}. Apart from the dice lattice, there are many other systems that host higher Chern numbers, such as, a decorated honeycomb structure or a star lattice \cite{chen2012}, a longer range hopping in the multiorbital triangular lattice \cite{sarma2012}, Dirac \cite{sticlet2013} and semi-Dirac \cite{mondal2022_1} systems etc. In presence of a spin-orbit coupling, a honeycomb lattice \cite{yang2014, yang2016}, and ultracold gases in a triangular lattice \cite{alase2021, goldman2021} also depict large Chern numbers. Further, in sonic crystals created using acoustic components \cite{zhao2022}, Cr-doped laminar sheets of $\mathrm{Bi}_2(\mathrm{Se, Te})_3$ \cite{zhang2013}, or a magnetic doped topological insulator \cite{bernevig2014}, higher values of Chern numbers are predicted. Experimentally, a multilayered structure consisting of alternatively arranged doped (with magnetic materials) and undoped topological insulator layers also reveals presence of several higher Chern numbers \cite{samarth2020} and the Hall conductivity is found to scale with the number of layers. Also $\mathrm{MnBi}_2\mathrm{Te}_4$ devices at high temperature show a Chern number, $C = 2$ \cite{ge2020, zhu2022}.

	In this paper we focus on a dice lattice that includes the TRS breaking complex NNN hopping term. The topological and the transport properties of such a system may be predictable. Owing to an additional flat band at the Fermi level, the topological properties are characterized by Chern numbers $\pm2$. Further, the electrons in the flat band can not move, and hence their contribution to the transport phenomena may not yield interesting results. However our studies will have larger impact if the flat band is rendered dispersive, which will enable it to contribute to the transport and may have important consequences on the topological properties. 
	We simulate the evolution of the topological properties of a band deformed dice lattice where the effects of deformation of the band structure are induced by an anisotropic hopping energy between selected nearest neighbour bonds. In a first attempt (we shall call this as case-I later), the hopping anisotropy is introduced via tuning the nearest nighbour (NN) hopping amplitudes between the A-B (say, $t_1$) and B-C (say, $t^\prime$) sublattices identically along a particular direction, while those along the rest of the NN directions ($t$) are kept unaltered (see Fig. \ref{dice_lattice}). Such an anisotropic hopping causes the band extrema from the Dirac points to move closer to each other, and they eventually merge at an intermediate $\mathbf{M}$ point. The spectral gaps and the topological properties vanish at a special value of the hopping $t_1$, namely, $t_1 = t^\prime = 2t$. In the absence of the NNN hopping, the band dispersion for this particular value of $t_1$ and $t^\prime$  is quadratic along the $k_x$ direction and linear along the $k_y$ direction, and is commonly referred to as the semi-Dirac dispersion. However, the presence of NNN hopping makes the spectrum anisotropic linear, that is, linear along both the directions but have unequal velocities. This phenomenon is similar to the case for graphene, where a topological phase transition takes place at the gap closing semi-Dirac point \cite{mondal2021,mondal2022_2}, except that in the dice lattice we have an additional zero energy flat band. Thus, disappearance of a band gap is more involved here as we shall see below.
	
	An alternative option to induce hopping anisotropy is via `\textit{selectively}' tuning the hopping $t_1$ between the A and B sublattices, while keeping $t^\prime$ between the B and C sublattices unaltered. To distinguish it from the case above, we call it case-II. Such a selective hopping anisotropy leads to distinct effects compared to the above case, where the band extrema do not migrate, however the flat band gets deformed, which eventually alters the gap between the conduction (or valence) and the flat band. Moreover, the deformation of the flat band (which does not remain flat any longer) imparts a dispersive character, and hence should enable it to contribute to the transport properties. Finally, the spectral gap vanishes at certain value of the anisotropy, namely, $t_1 = 1.67t$ which is different from case-I (that is, $t_1=t^\prime=2t$) and depends on the value of NNN hopping amplitude $t_2$. The latter is also a discernible feature of this system, since the value of $t_2$ did not play a role earlier and any non-zero $t_2$ induces topological properties. Such vanishing of the band gap again leads to the vanishing of the topological properties similar to the case above. However, the Chern insulating regions in the phase diagram and the anomalous Hall conductivities have differences in these two cases. Specifically, the effects of the flat band being dispersive will lead to observable consequences in the behaviour of the anomalous Hall conductivity.

	Possible experimental scenarios in inducing hopping anisotropy, at least for case-I can be provided as follows.
	The hopping energies $t_1$ and $t^\prime$ can be altered simultaneously by applying an uniaxial strain to the system which changes the bond length along the direction of application of the strain. Thus, the hopping energies among the A-B and B-C sublattices  will be modified simultaneously. Such applications of uniaxial strain in the honeycomb lattice structure, such as, $\mathrm{Si_2O}$ yields a semi-Dirac band structure \cite{zhong2017}. For the sake of completeness, we have applied both type of anisotropy to see the properties of the dice lattice. Case-II involves selective control of the hopping amplitudes (see Fig. \ref{dice_lattice}) and is included for comparison owing to the interesting consequences detailed above (and also later in section \ref{sec:topological_properties}). In general, the semi-Dirac dispersion (albeit without the Haldane flux) is somewhat eastablished in experiments, such as, the $\mathrm{BEDT}$-$\mathrm{TTF_2I_3}$ organic salts under pressure \cite{suzumura2013,hasegawa2006}, in multilayered structures of $\mathrm{TiO}_2/\mathrm{VO}_2$ \cite{pickett2009,pickett2010}, black phosphorene doped by means of \textit{in situ}  deposition of potassium atoms \cite{kim2015} etc.

	The paper is organized in the following way. In Sec. \ref{sec:model_hamiltonian}, we present the Hamiltonian of the system for the two cases (case-I and case-II) and the energy dispersions are plotted for different values of the anisotropy parameters in Sec. \ref{sec:bandstructure}. Sec. \ref{sec:topological_properties} deals with the topological properties in which Sec. \ref{sec:phase_diagram} shows the phase diagrams corresponding to different values of the anisotropic hoppings. Further, the presence (or absence) of a pair of edge modes of a nanoribbon for various parameters are shown in Sec. \ref{sec:edge_states}. Finally we present numerical computation of the anomalous Hall conductivity in Sec. \ref{sec:hall_conductivity} for the cases-I and II. We finally conclude with a brief summary of the results obtained in Sec. \ref{sec:conclusion}.
	
	\section{The Hamiltonian}\label{sec:model_hamiltonian}
	\begin{figure}[h]
		\centering
		\includegraphics[width=0.4\textwidth]{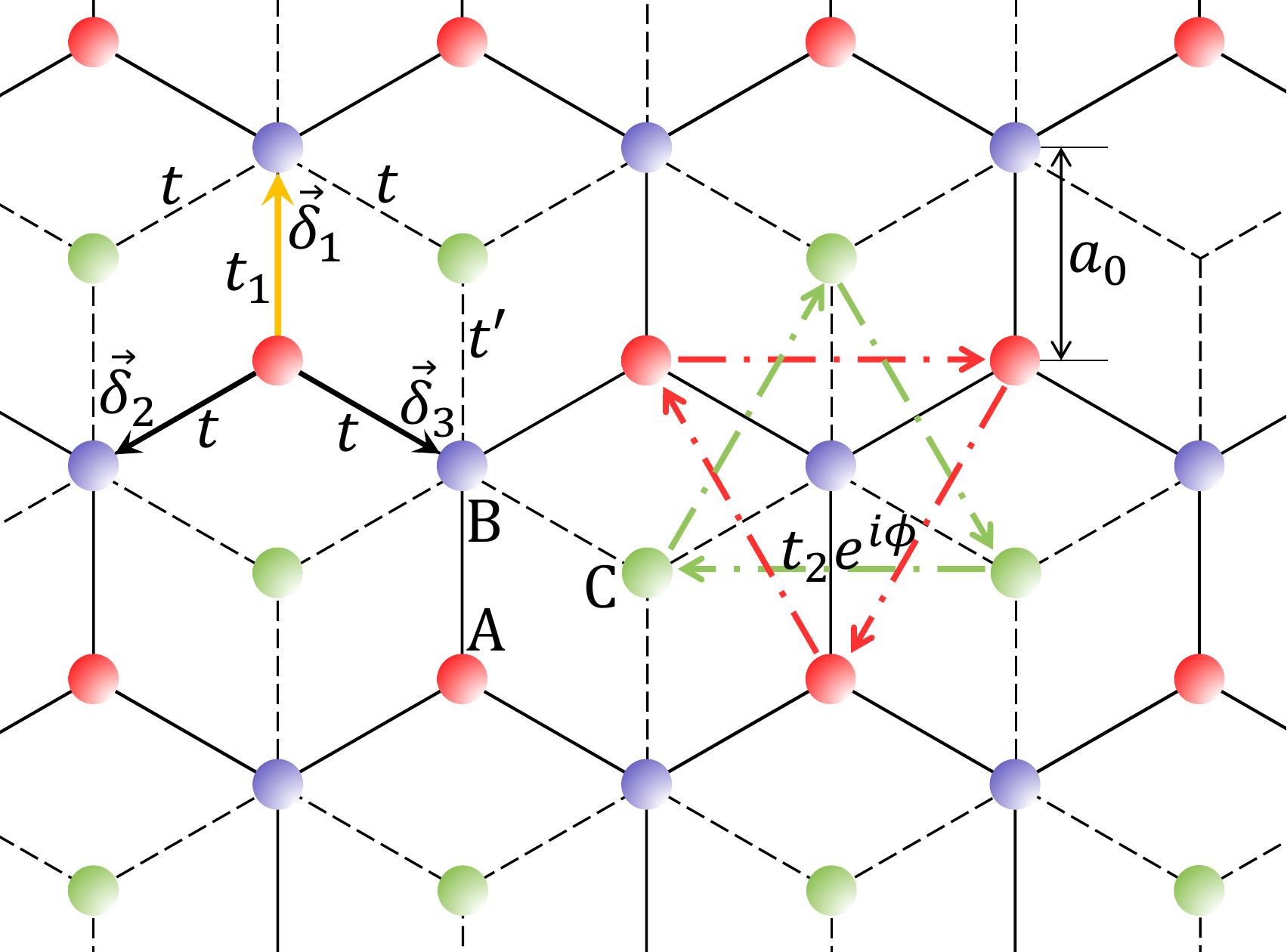}
		\caption{\raggedright A dice lattice is shown where the red, blue and the green circles represent the sublattices A, B and C respectively. The NN hopping strength between A and B sublattices along the $\boldsymbol{\delta}_1$ direction (shown via the yellow arrow) is $t_1$, while it is $t^\prime$ between B and C sublattices along the same direction. The NNN hopping is $t_2e^{i\phi}$ ($t_2e^{-i\phi}$) for the clockwise (anti-clockwise) hopping direction. $\boldsymbol{\delta}_i$s and $a_0$ represent the NN vectors and lattice constant respectively.}\label{dice_lattice}
	\end{figure}
	\begin{figure}[h]
		\captionsetup[subfigure]{labelformat=nocaption}
		\centering
		\includegraphics[width=\linewidth]{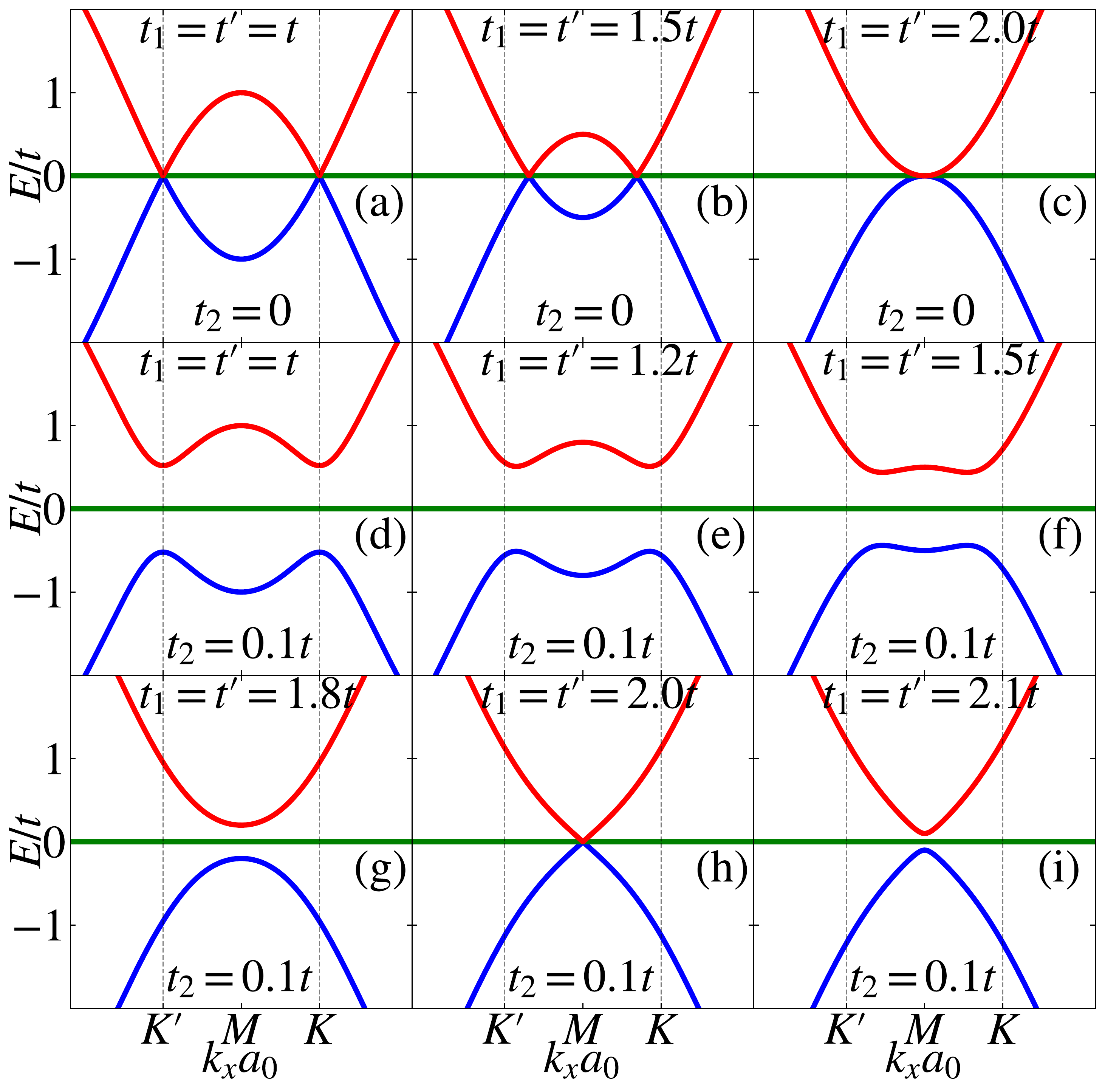}
		
		\begin{subfigure}[b]{0\textwidth}
			\subcaption{}\label{fig:band_t1_10_t2_0_m_0_ABC}
		\end{subfigure}
		\begin{subfigure}[b]{0\textwidth}
			\subcaption{}\label{fig:band_t1_11_t2_0_m_0_ABC}
		\end{subfigure}
		\begin{subfigure}[b]{0\textwidth}
			\subcaption{}\label{fig:band_t1_12_t2_0_m_0_ABC}
		\end{subfigure}
		\begin{subfigure}[b]{0\textwidth}
			\subcaption{}\label{fig:band_t1_10_t2_100_m_0_ABC}
		\end{subfigure}
		\begin{subfigure}[b]{0\textwidth}
			\subcaption{}\label{fig:band_t1_11_t2_100_m_0_ABC}
		\end{subfigure}
		\begin{subfigure}[b]{0\textwidth}
			\subcaption{}\label{fig:band_t1_12_t2_100_m_0_ABC}
		\end{subfigure}
		\begin{subfigure}[b]{0.\textwidth}
			\subcaption{}\label{fig:band_t1_13_t2_100_m_0_ABC}
		\end{subfigure}
		\begin{subfigure}[b]{0.\textwidth}
			\subcaption{}\label{fig:band_t1_14_t2_100_m_0_ABC}
		\end{subfigure}
		\begin{subfigure}[b]{0.\textwidth}
			\subcaption{}\label{fig:band_t1_15_t2_100_m_0_ABC}
		\end{subfigure}
		\caption{\raggedright The band structure of the system in absence of $t_2$ $(t_2=0)$ is shown along the $k_x$-axis  (at $k_ya_0 = 2\pi/3$) for (a) $t_1 = t^\prime = t$, (b) $t_1 = t^\prime = 1.5t$ and (c) $t_1 = t^\prime = 2t$. Similarly, the dispersion  in presence of $t_2$ $(t_2=0.1t)$ is depicted for
		(d) $t_1 = t^\prime = t$, (e) $t_1 = t^\prime = 1.2t$, (f) $t_1 = t^\prime = 1.5t$, (g) $t_1 = t^\prime = 1.8t$, (h) $t_1 = t^\prime = 2t$, and (i) $t_1 = t^\prime = 2.1t$. In the figures, $k_x$ is rendered dimensionless by multiplying with lattice constant $a_0$.}
		\label{fig:band1}
	\end{figure}

	A tight-binding Hamiltonian on a dice lattice can be written as follows,
	\begin{align}\label{eq:ham1}
		H = & \left[ \sum_{\langle ij \rangle} t_{ij} c_i^\dagger c_j + \sum_{\langle\langle ij \rangle\rangle} e^{i\phi_{ij}} c_i^\dagger c_j  + \mathrm{h.c.} \right] +  \sum_i \Delta_i c_i^\dagger c_i
	\end{align}
	The first term is the nearest neighbour (NN) hopping, where the hopping $t_{ij} = t_1$ when $i$ connects the site $j$ along the $\boldsymbol{\delta}_1 = a_0 (0, 1)$ direction, while it is $t$ along the $\boldsymbol{\delta}_2= a_0(\sqrt{3}/2, -1/2)$ and $\boldsymbol{\delta}_3= a_0(\sqrt{3}/2, -1/2)$ directions shown in Fig. \ref{dice_lattice}. 
	Such values of $t_{ij}$ are restricted among the A and B sublattices. Further, the NN hoppings between the B and C sublattices along the $\boldsymbol{\delta}_1$ direction is $t^\prime$ and $t$ in the $\boldsymbol{\delta}_{2, 3}$ directions. As detailed earlier, we have considered two cases. In case-I we vary both $t_1$ and $t^\prime$ in the range $[t:2t]$, while is the second case, $t^\prime$ is kept fixed at $t$, and $t_1$ is varied in the range $[t:1.8t]$. 
	The second term in Eq. \ref{eq:ham1} represents the complex next nearest neighbour hopping (NNN) with an amplitude $t_2$ and a phase $\phi_{ij}$, where $\phi_{ij}$ is positive (negative) when an electron hops in the clockwise (anti-clockwise) direction. The third term denotes the onsite energy term (Semenoff mass), that assumes values $\pm\Delta$ at the sites that belong to B and C sublattices respectively. 
	
	The Fourier transformed Hamiltonian can be written as,
	\begin{equation}\label{eq:ham_kspace}
		H(\mathbf{k}) = 
		\begin{pmatrix}
			h_z(\mathbf{k}) & h(\mathbf{k}, t_1) & 0 \\
			h^*(\mathbf{k},  t_1) & 0 & h(\mathbf{k}, t^\prime)\\
			0 & h^*(\mathbf{k}, t^\prime) & -h_z(\mathbf{k})
		\end{pmatrix}
	\end{equation}
	where $h(\mathbf{k}, \tilde{t}) = h_x(\mathbf{k}, \tilde{t}) - i h_y(\mathbf{k}, \tilde{t})$ with $\tilde{t}$ being either $t_1$ or $t^\prime$. The expressions for the  $h_i$s can be written as,
	\begin{equation}
		h_x(\mathbf{k}, \tilde{t}) = \{\tilde{t} \cos k_y + 2t\cos \frac{k_y}{2} \cos\frac{\sqrt{3}k_x}{2}\},
	\end{equation}
	\begin{equation}
		h_y(\mathbf{k}, \tilde{t}) = \{-\tilde{t} \sin k_y + 2t\sin \frac{k_y}{2} \cos\frac{\sqrt{3}k_x}{2}\}
	\end{equation}
	and,
	\begin{align}
		h_z(\mathbf{k})= \Delta-2t_2\sin \phi\left\{2\sin\frac{\sqrt{3}k_x}{2} \cos\frac{3k_y}{2} - \sin\sqrt{3}k_x \right\}
	\end{align}
	
	
	The Haldane flux is kept constant at $\pi/2$ so as to render the NNN hopping purely imaginary, and we have considered $\Delta=0$, except for the computation of the phase diagram in Sec. \ref{sec:phase_diagram}.

\section{Spectral properties}\label{sec:bandstructure}
	The electronic energy spectra of the system, where both $t_1$ and $t^\prime$ are varied together, have been obtained via numerical diagonalization of Eq. \ref{eq:ham_kspace} and are shown in Fig. \ref{fig:band1}. Three bands appear in the spectrum, which we term as the conduction band (shown in red), flat band (in green) and valence band (in blue). In absence of the NNN hopping, there is no spectral gap anywhere (Figs. \ref{fig:band_t1_10_t2_0_m_0_ABC}-\ref{fig:band_t1_12_t2_0_m_0_ABC}) in the BZ. As we turn on the NNN hopping, spectral gaps (of same magnitude) open up at the $\mathbf{K}\,(2\pi/3\sqrt{3}a_0, 2\pi/3a_0)$ and $\mathbf{K^\prime}\,(-2\pi/3\sqrt{3}a_0, 2\pi/3a_0)$ points (Fig. \ref{fig:band_t1_10_t2_100_m_0_ABC}). Now if we increase $t_1$ and $t^\prime$, the band extrema at the $\mathbf{K}$ and the $\mathbf{K}^\prime$ points migrate towards each other which results in diminishing of the band gap. At a special value of the hopping amplitude, namely, $t_1 =t^\prime= 2t$, the band gap vanishes at $\mathbf{M} \,(2\pi/3a_0, 0)$ point even in the presence of the Haldane term, $t_2$ (tat is, TRS remains broken) (Fig. \ref{fig:band_t1_14_t2_100_m_0_ABC}). As both $t_1$ and $t^\prime$ are increased beyond value $2t$ ($t$ denotes NN hopping), a spectral gap opens up again at the $\mathbf{M}$ point and the band structure henceforth remains gapped for all values of $t_1>2t$ and $t^\prime>2t$. It should be noted that without the Haldane's NNN hopping, the band structure of the dice lattice (Figs. \ref{fig:band_t1_10_t2_0_m_0_ABC}-\ref{fig:band_t1_12_t2_0_m_0_ABC}) demonstrates very similar properties with that of graphene, except that there is no flat band for the latter \cite{dietl2008, banerjee2009}. At $t_1 = t^\prime = 2t$, the spectrum resembles a semi-Dirac dispersion, that is, linear along  $k_y$, and quadratic along the $k_x$ direction. The presence of $t_2$ makes the dispersion anisotropic linear, that is linear along both the directions, however the electrons move with different velocities (see Fig. \ref{fig:band_t1_14_t2_100_m_0_ABC}). Further, the spectra for other values of $t_1$ (for a finite $t_2$) demonstrate similar features as that obtained for graphene \cite{mondal2021}. 
	
\begin{figure}[h]
		\captionsetup[subfigure]{labelformat=nocaption}
		\centering
		\includegraphics[width=\linewidth]{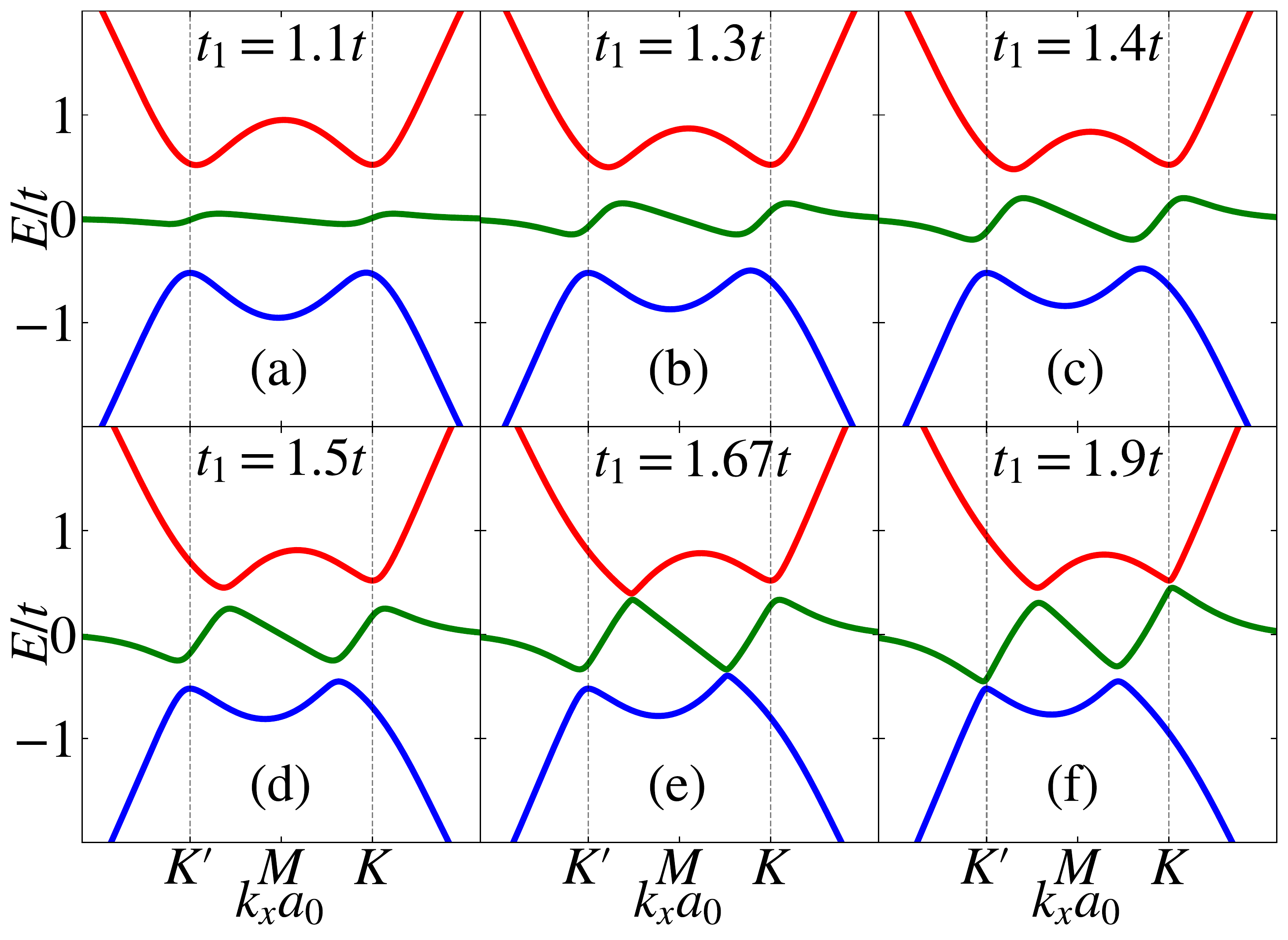}
		\begin{subfigure}[b]{0\textwidth}
			\subcaption{}\label{fig:band_t1_10_t2_100_m_0}
		\end{subfigure}
		\begin{subfigure}[b]{0\textwidth}
			\subcaption{}\label{fig:band_t1_11_t2_100_m_0}
		\end{subfigure}
		\begin{subfigure}[b]{0\textwidth}
			\subcaption{}\label{fig:band_t1_12_t2_100_m_0}
		\end{subfigure}
		\begin{subfigure}[b]{0.\textwidth}
			\subcaption{}\label{fig:band_t1_13_t2_100_m_0}
		\end{subfigure}
		\begin{subfigure}[b]{0.\textwidth}
			\subcaption{}\label{fig:band_t1_14_t2_100_m_0}
		\end{subfigure}
		\begin{subfigure}[b]{0.\textwidth}
			\subcaption{}\label{fig:band_t1_15_t2_100_m_0}
		\end{subfigure}
		\caption{\raggedright The band structure of the system is shown along the dimensionless $k_x$-axis  (at $k_ya_0 = 2\pi/3$) for (a) $t_1 = 1.1t$, (b) $t_1 = 1.3t$, (c) $t_1 = 1.4t$, (d) $t_1 = 1.5t$, (e) $t_1 = 1.67$, and (f) $t_1 = 1.9$.  The values of $t^\prime$, $t_2$, $\phi$ and $\Delta$ are taken as $t$, $0.1t$, $\pi/2$ and zero respectively.}
		\label{fig:band2}
	\end{figure}

	Now we discuss the spectral features corresponding to case-II where $t^\prime$ is fixed at $t$, while $t_1$ is varied. The corresponding plots are depicted in Fig. \ref{fig:band2}.
	As can be seen the flat band becomes dispersive for $t_1 \neq t$, and the spectral gap decreases. Hence, we can no longer call it flat band as it becomes dispersive and we refer to it from now on as the middle band. It should be noted that the conduction band minimum remains fixed at the $\mathbf{K}$ point, while the same near the $\mathbf{K^\prime}$ point is displaced along the positive $k_x$-direction. A similar scenario occurs in the case of the valence band. In fact the reverse occurs, that is, the valence band maximum at $\mathbf{K^\prime}$ point remains constant, while that at $\mathbf{K}$ shifts. Finally the gap vanishes completely at a specific value of $t_1$, namely $t_1 \simeq 1.67t$. Beyond $t_1 \simeq 1.67t$, the gap reopens. Thus the gap closing or the so called `\textit{semi-Dirac}' limit occurs at much lower value compared to $t_1 = 2t$. Another important aspect of keeping $t^\prime$ unchanged ($t^\prime = t$) is that the closing of the energy gap depends on the value of $t_2$. The spectrum shown in Fig. \ref{fig:band2} is for $t_2 = 0.1t$ for which the gap closes at $t_1 = 1.67t$. However, if $t_2$ is increases (decreases) the gap closes at higher (lower) values of $t_1$. Further, as opposed to all the three bands (three fold degeneracy) touching together at the $\mathbf{M}$ point in case-I, here we observed two bands (conduction and middle bands) touching each other above the Fermi level, while the other two (valence and middle bands) touch below the Fermi level (see Fig. \ref{fig:band_t1_14_t2_100_m_0}).
	

\section{Topological properties}\label{sec:topological_properties}	
\subsection{Chern number}\label{sec:phase_diagram}

\begin{figure*}[!htb]
	\captionsetup[subfigure]{labelformat=nocaption}
	\centering
	\includegraphics[width=0.32\linewidth]{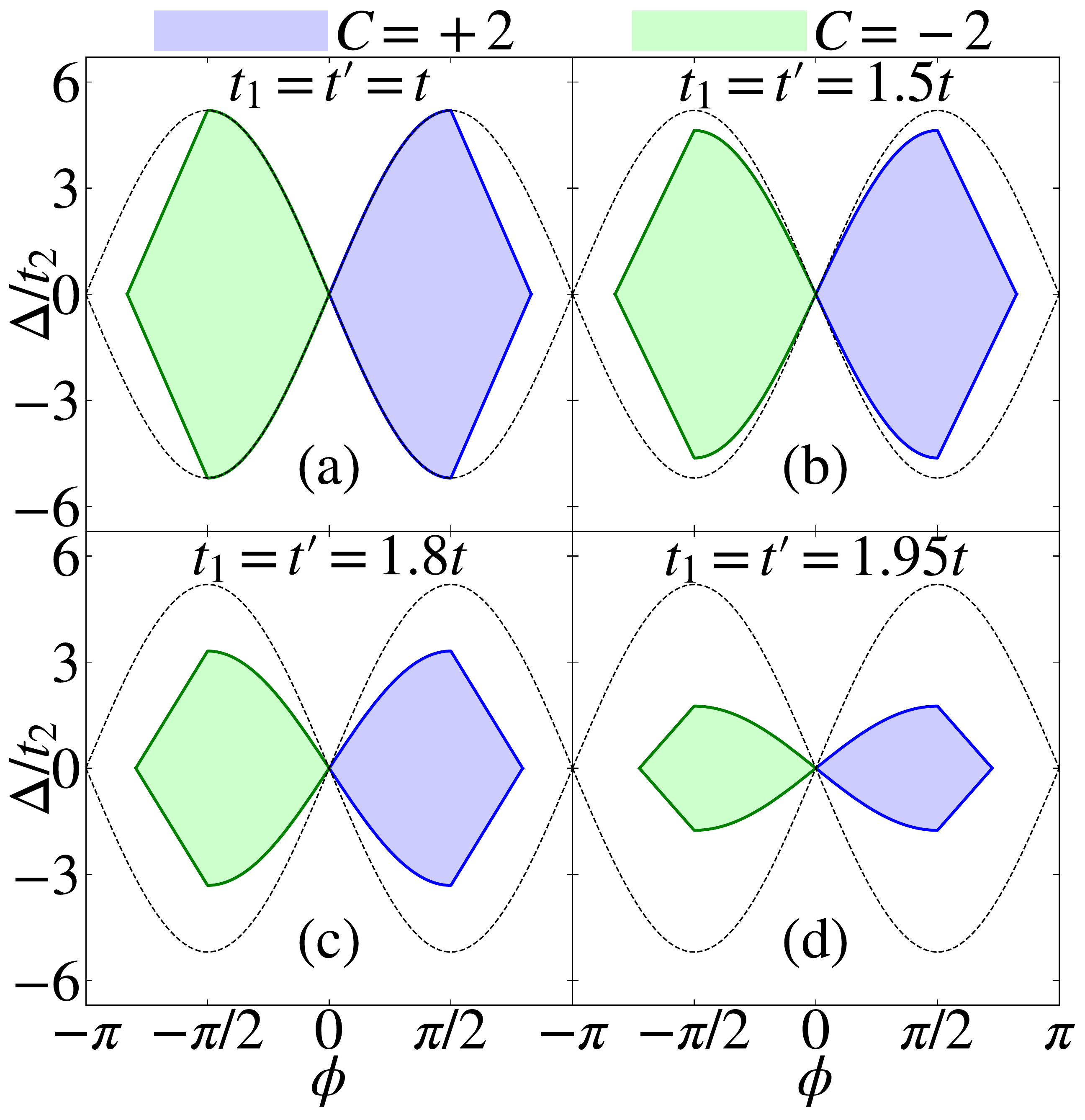}
	\includegraphics[width=0.32\linewidth]{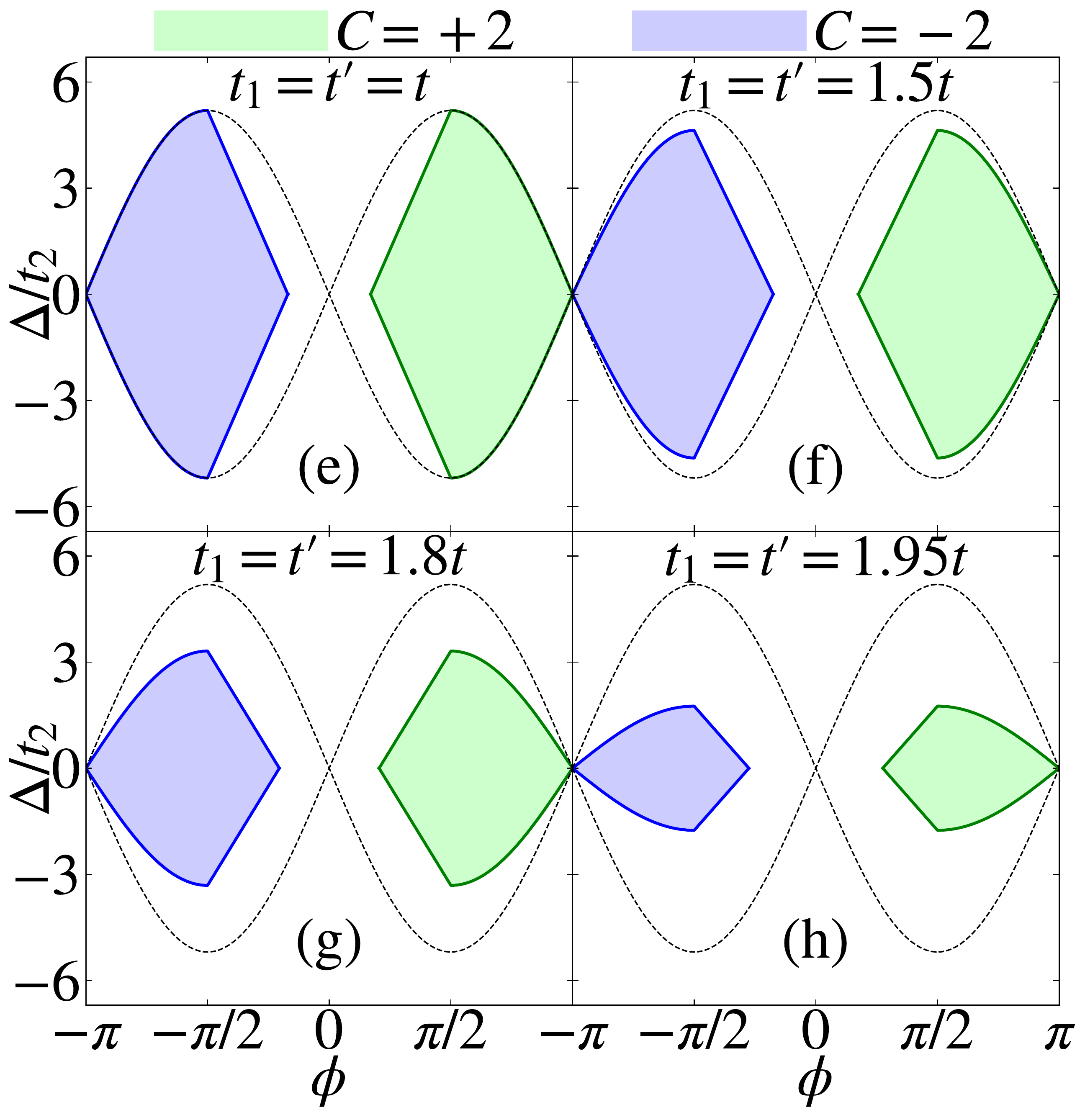}
	\includegraphics[width=0.32\linewidth]{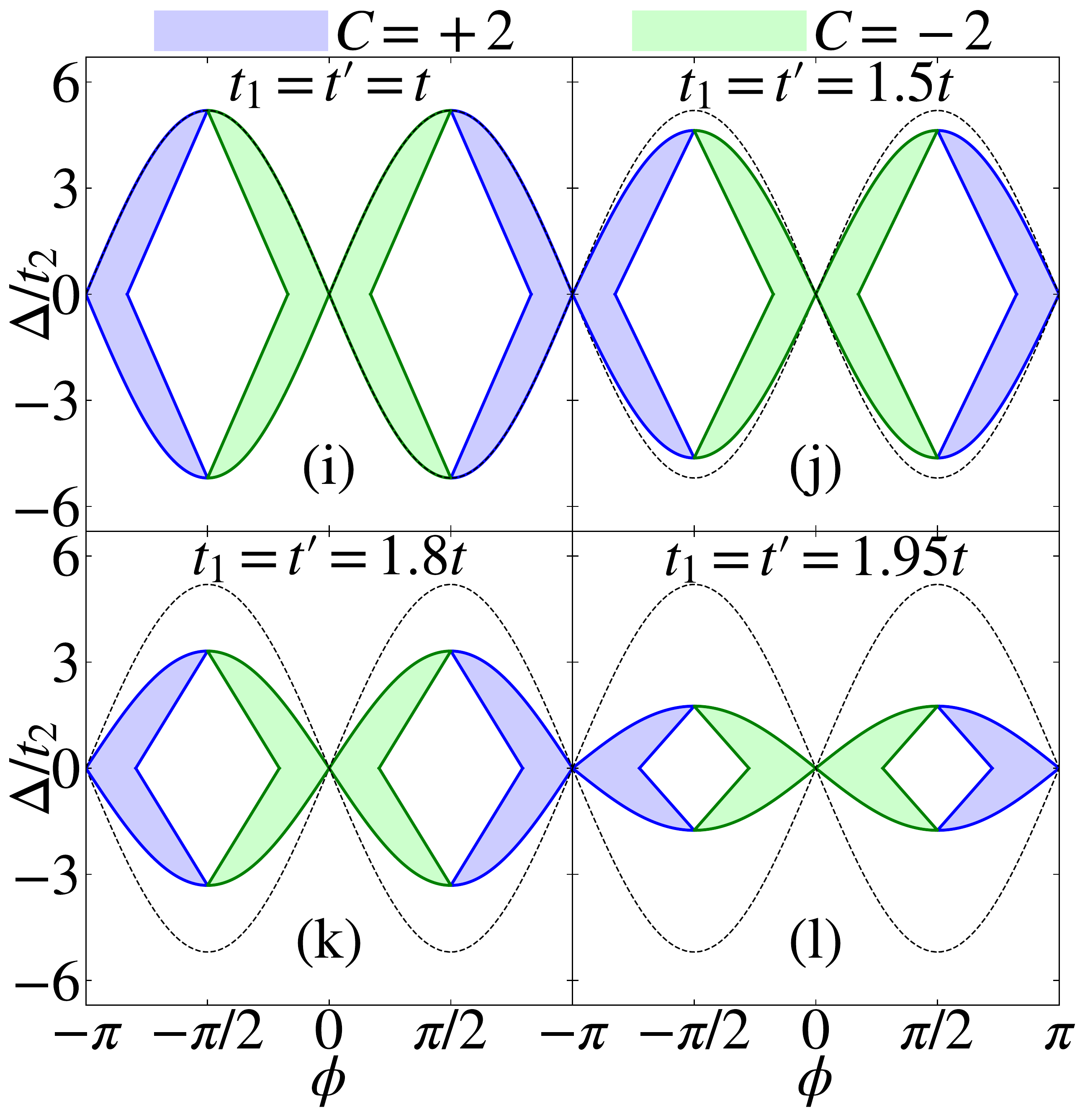}
	\begin{subfigure}[b]{0\textwidth}
		\subcaption{}\label{fig:pd_t1_1.0_t2_100_ABC_lb}
	\end{subfigure}
	\begin{subfigure}[b]{0\textwidth}
		\subcaption{}\label{fig:pd_t1_1.5_t2_0.1_ABC_lb}
	\end{subfigure}
	\begin{subfigure}[b]{0\textwidth}
		\subcaption{}\label{fig:pd_t1_1.8_t2_0.1_ABC_lb}
	\end{subfigure}
	\begin{subfigure}[b]{0.\textwidth}
		\subcaption{}\label{fig:pd_t1_1.95_t2_0.1_ABC_lb}
	\end{subfigure}
	\begin{subfigure}[b]{0\textwidth}
		\subcaption{}\label{fig:pd_t1_1.0_t2_100_ABC_ub}
	\end{subfigure}
	\begin{subfigure}[b]{0\textwidth}
		\subcaption{}\label{fig:pd_t1_1.5_t2_0.1_ABC_ub}
	\end{subfigure}
	\begin{subfigure}[b]{0\textwidth}
		\subcaption{}\label{fig:pd_t1_1.8_t2_0.1_ABC_ub}
	\end{subfigure}
	\begin{subfigure}[b]{0.\textwidth}
		\subcaption{}\label{fig:pd_t1_1.95_t2_0.1_ABC_ub}
	\end{subfigure}
	\begin{subfigure}[b]{0\textwidth}
		\subcaption{}\label{fig:pd_t1_1.0_t2_0.1_ABC_mb}
	\end{subfigure}
	\begin{subfigure}[b]{0\textwidth}
		\subcaption{}\label{fig:pd_t1_1.5_t2_0.1_ABC_mb}
	\end{subfigure}
	\begin{subfigure}[b]{0\textwidth}
		\subcaption{}\label{fig:pd_t1_1.8_t2_0.1_ABC_mb}
	\end{subfigure}
	\begin{subfigure}[b]{0.\textwidth}
		\subcaption{}\label{fig:pd_t1_1.95_t2_0.1_ABC_mb}
	\end{subfigure}
	\caption{\raggedright The phase diagrams are shown for $t_1 = t^\prime = t$ in (a), (e) and (i), $t_1 = t^\prime = 1.5t$ in (b), (f) and (j), $t_1 = t^\prime = 1.8t$ in (c), (g) and (k), $t_1 = t^\prime = 1.95t$ in (d), (h) and (l). These phase diagrams are presented corresponding to the valence, conduction and flat bands in (a)-(d), (e)-(h) and (i)-(l) respectively. The non-zero Chern numbers corresponding to blue and green regions have values $+2$ and $-2$ respectively (indicated above the figures), while the white regions represent vanishing Chern number.}
	\label{fig:pd_lb_ub_mb_ABC}
\end{figure*}

\begin{figure*}[!htb]
	\captionsetup[subfigure]{labelformat=nocaption}
	\centering
	\includegraphics[width=0.32\linewidth]{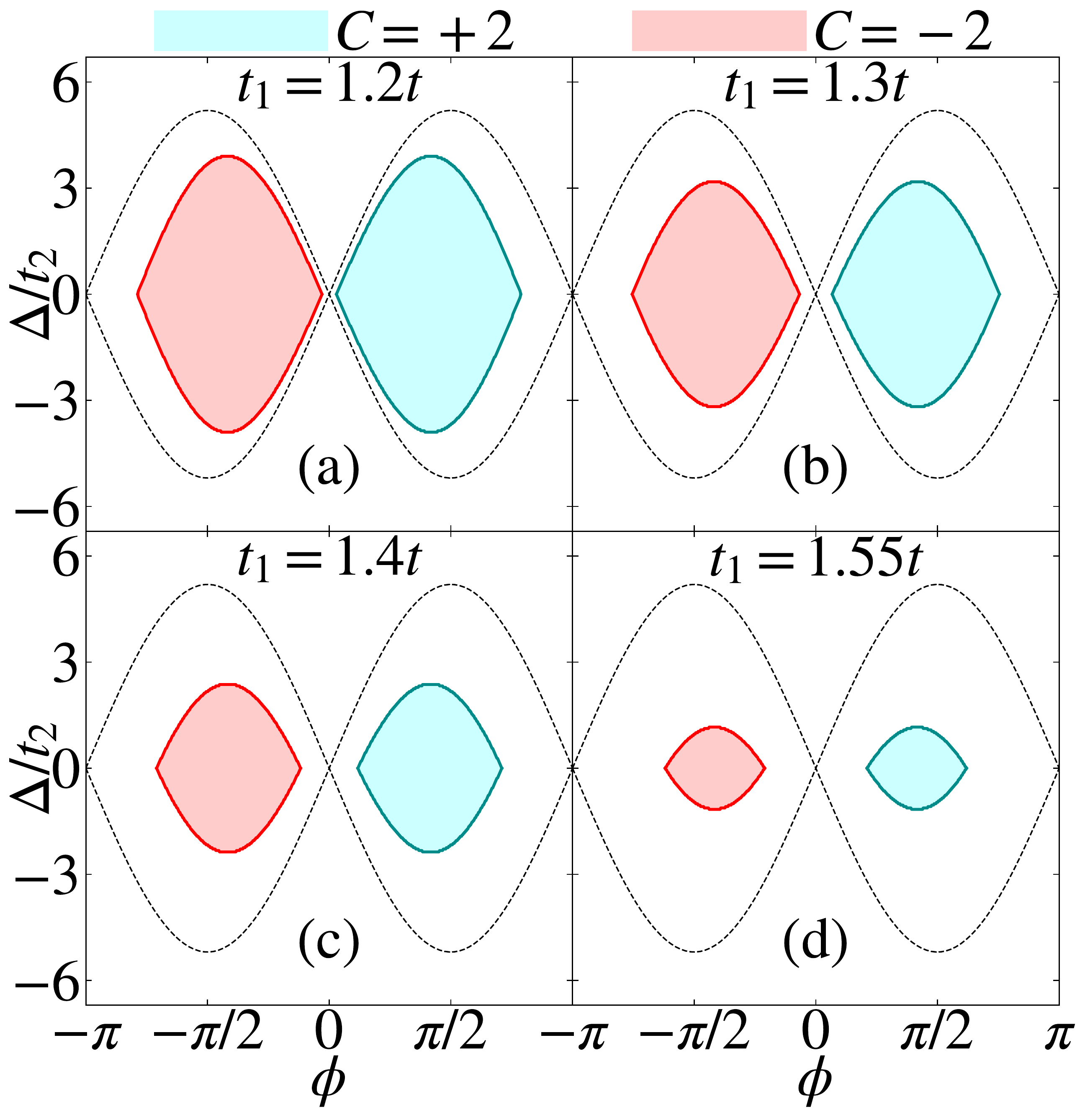}
	\includegraphics[width=0.32\linewidth]{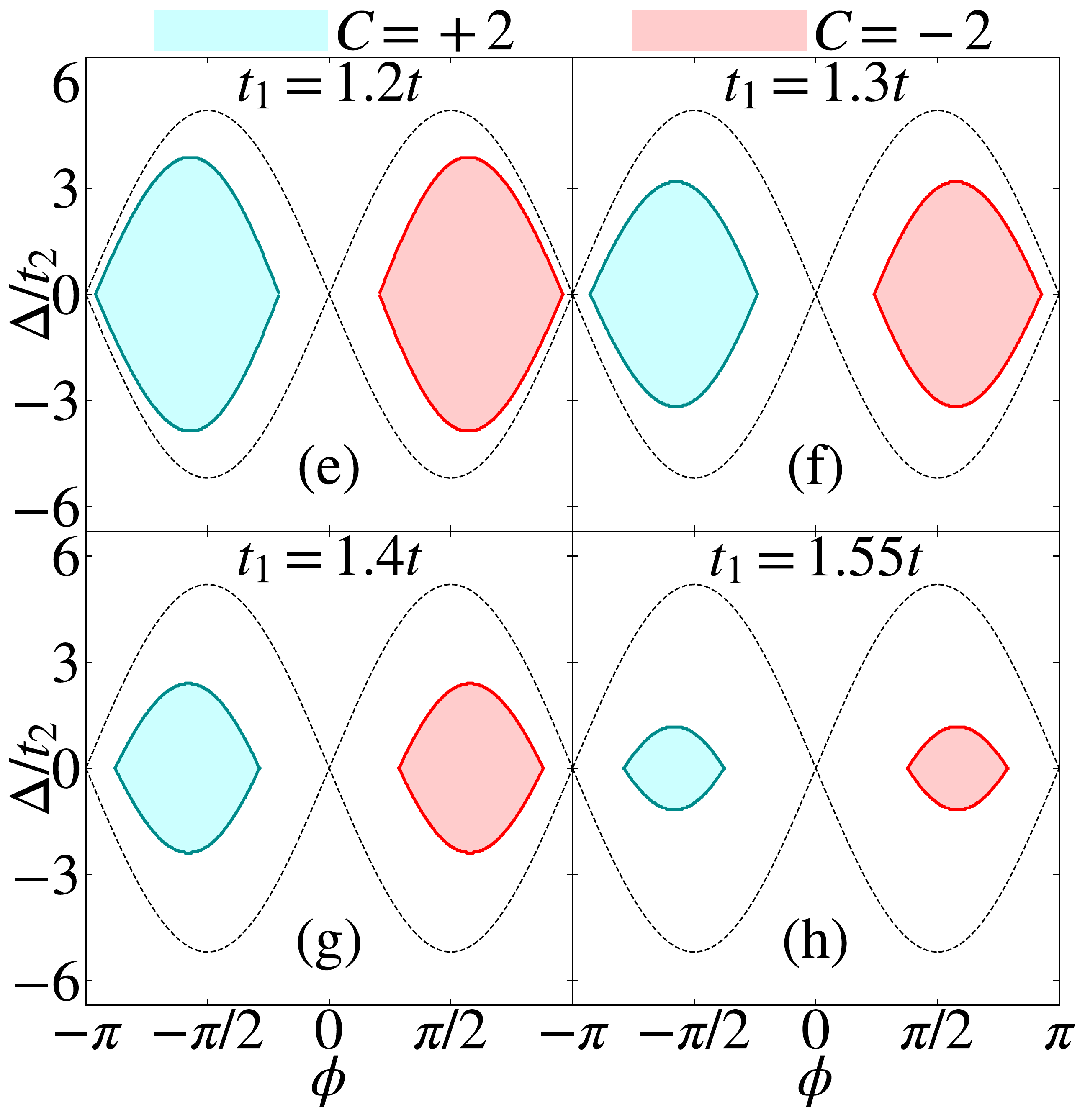}
	\includegraphics[width=0.32\linewidth]{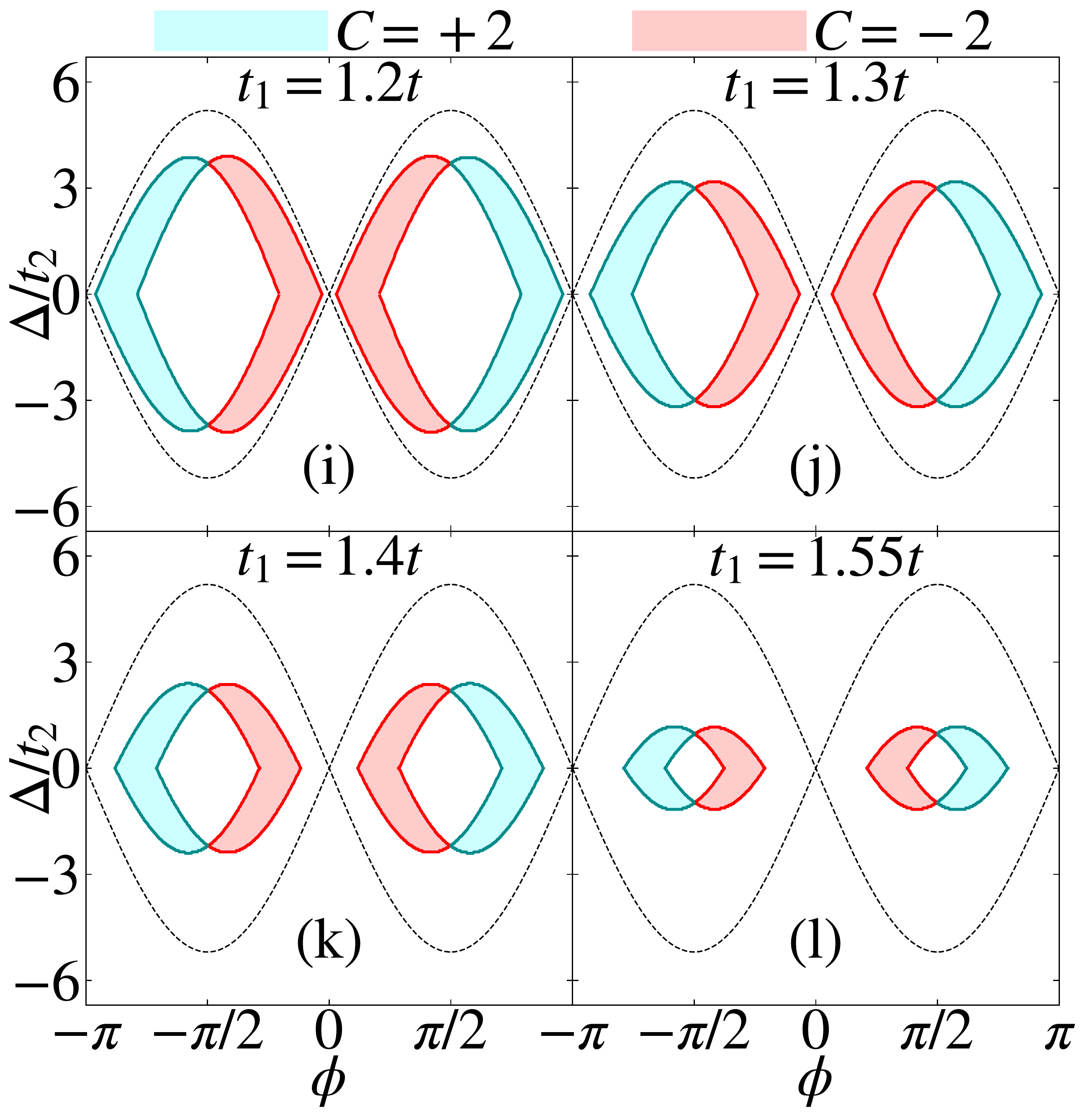}
	\begin{subfigure}[b]{0\textwidth}
		\subcaption{}\label{fig:pd_t1_1.2_t2_0.1_lb}
	\end{subfigure}
	\begin{subfigure}[b]{0\textwidth}
		\subcaption{}\label{fig:pd_t1_1.3_t2_0.1_lb}
	\end{subfigure}
	\begin{subfigure}[b]{0\textwidth}
		\subcaption{}\label{fig:pd_t1_1.4_t2_0.1_lb}
	\end{subfigure}
	\begin{subfigure}[b]{0.\textwidth}
		\subcaption{}\label{fig:pd_t1_1.55_t2_0.1_lb}
	\end{subfigure}
	\begin{subfigure}[b]{0\textwidth}
		\subcaption{}\label{fig:pd_t1_1.2_t2_0.1_ub}
	\end{subfigure}
	\begin{subfigure}[b]{0\textwidth}
		\subcaption{}\label{fig:pd_t1_1.3_t2_0.1_ub}
	\end{subfigure}
	\begin{subfigure}[b]{0\textwidth}
		\subcaption{}\label{fig:pd_t1_1.4_t2_0.1_ub}
	\end{subfigure}
	\begin{subfigure}[b]{0.\textwidth}
		\subcaption{}\label{fig:pd_t1_1.55_t2_0.1_ub}
	\end{subfigure}
	\begin{subfigure}[b]{0\textwidth}
		\subcaption{}\label{fig:pd_t1_1.2_t2_0.1_mb}
	\end{subfigure}
	\begin{subfigure}[b]{0\textwidth}
		\subcaption{}\label{fig:pd_t1_1.3_t2_0.1_mb}
	\end{subfigure}
	\begin{subfigure}[b]{0\textwidth}
		\subcaption{}\label{fig:pd_t1_1.4_t2_0.1_mb}
	\end{subfigure}
	\begin{subfigure}[b]{0.\textwidth}
		\subcaption{}\label{fig:pd_t1_1.55_t2_0.1_mb}
	\end{subfigure}
	\caption{\raggedright The phase diagrams are shown for $t_1 = 1.2t$ in (a), (e) and (i), $t_1 = 1.3t$ in (b), (f) and (j), $t_1 = 1.4t$ in (c), (g) and (k), $t_1 = 1.55t$ in (d), (h) and (l). These phase diagrams are presented corresponding to the valence, conduction and the middle bands in (a)-(d), (e)-(h) and (i)-(l) respectively. The value of  $t^\prime$ and $t_2$ are taken as $t$ and $0.1t$ respectively. The non-zero Chern numbers corresponding to cyan and red regions have values $+2$ and $-2$ respectively (indicated above the figures), while the white regions represent vanishing Chern number.}
	\label{fig:pd_lb_ub_mb}
\end{figure*}

\begin{figure}[h]
	\centering
	\includegraphics[width=\linewidth]{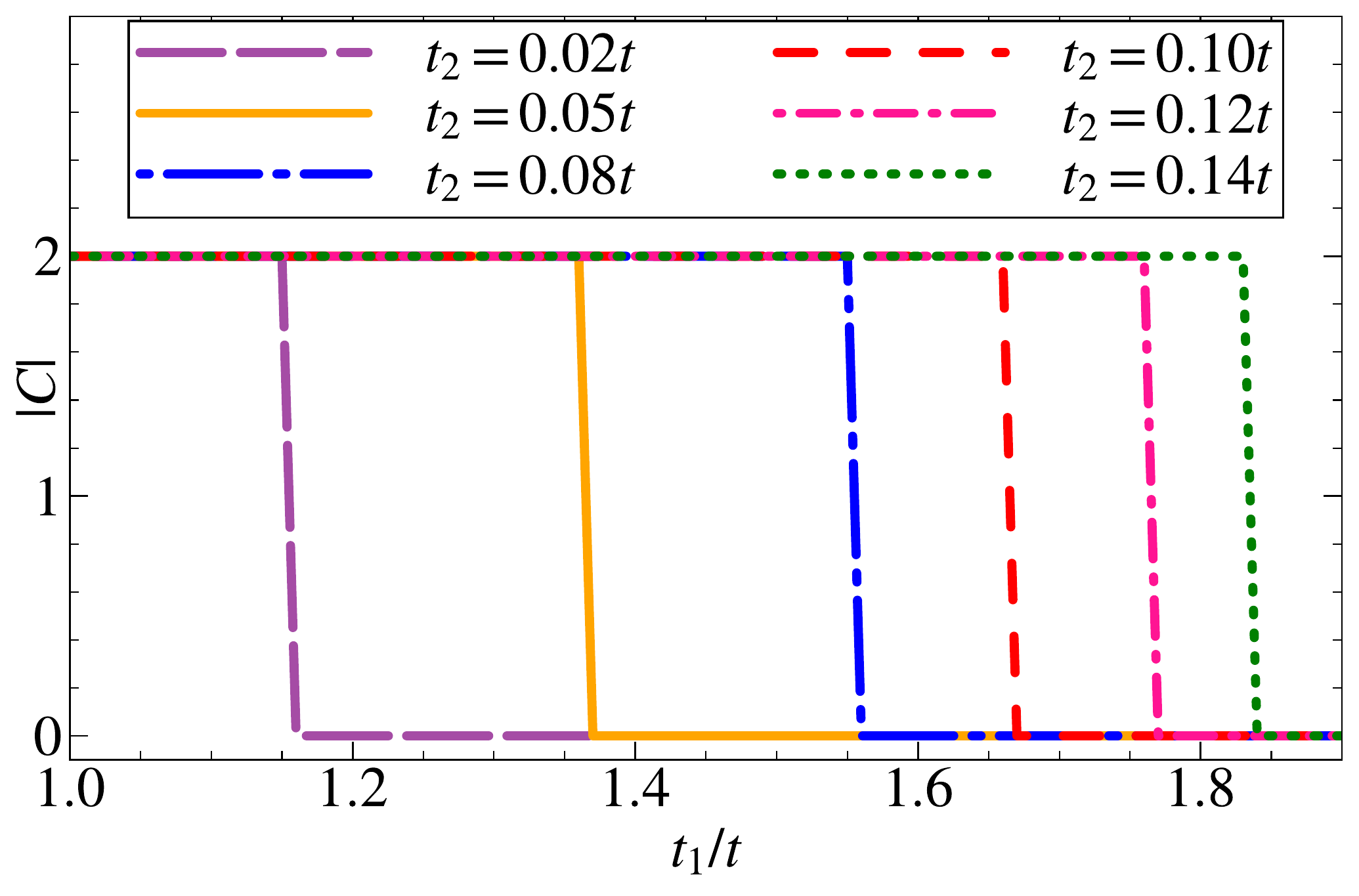}
	\caption{\raggedright The variation of Chern number as a function of the NN hopping amplitude $t_1/t$ is shown for various values of $t_2$, which are indicated in panel above the curves.}\label{fig:C_vs_t1}
\end{figure}

	In this section, we obtain the phase diagram of the system by calculating the Chern number numerically. Since the NNN hopping term breaks the TRS, we should get the non-zero values of Chern number. Moreover, finite values of the onsite energy also play an important role in opening or closing of the energy gaps at the Dirac points which play an essential role in inducing a topological phase transition. The Chern number ($C$) of a system can be calculated from the relation \cite{thouless, avron1988},
	\begin{eqnarray}\label{eq:chern_number}
		C & = & \frac{1}{2\pi}\int\int_{\mathrm{BZ}}\Omega(k_x, k_y)\mathrm{d}k_x \mathrm{d}k_y
	\end{eqnarray}
	Here $\Omega(k_x, k_y)$ is the $z$-component of the Berry curvature \cite{liu2016} which is given by,
	\begin{eqnarray}\label{eq:berry_curv}
		\Omega(k_x, k_y) = -2i\mathrm{Im}\left[\left< \frac{\partial\psi(k_x, k_y)}{\partial k_x} \right.\left|\frac{\partial\psi(k_x, k_y)}{\partial k_y}\right>\right]
	\end{eqnarray} 
	where $\psi(k_x, k_y)$ is the Hamiltonian defined in Eq. \ref{eq:ham_kspace} and $\mathrm{Im}$ denotes the imaginary part. Hence we obtain the Chern numbers by varying both $t_1$ and $t^\prime$ (case-I) as a function of $\Delta$ and $\phi$ as presented in the phase diagrams in Fig. \ref{fig:pd_lb_ub_mb_ABC}. As can be noticed,
	the green and the blue regions in each figure represent the topological phases of the system with Chern numbers $C=+2$ and $-2$ respectively, while the white region denotes trivial phase with zero Chern number ($C = 0$). Also, we have shown the boundaries separating the topological and the trivial phases corresponding to the original Haldane model for graphene by the black sinusoidal curve. When $t_1 = t^\prime = t$, that is, for the familiar dice lattice we get a maximum area of Chern insulating lobes (Fig. \ref{fig:pd_t1_1.0_t2_100_ABC_lb}). However, the Chern insulating (topological) regions are smaller than those for the original Haldane model. The phase boundary is sinusoidal for $0\leq|\phi|\leq \pi/2$ and linear for $|\phi| > \pi/2$.   Further, with the increase in the value of $t_1$ and $t^\prime$ (Figs. \ref{fig:pd_t1_1.5_t2_0.1_ABC_lb}-\ref{fig:pd_t1_1.95_t2_0.1_ABC_lb}), the area of the Chern insulating region gradually decreases, but the phase boundary follows the pattern corresponding to that of $t_1 = t^\prime = t$. Finally, the Chern number vanishes completely at the semi-Dirac limit, that is, $t_1=t^\prime= 2t$. For any non-zero value of $\Delta$, there is always a gap in the band structure, however $|C|$ remains zero. 
	If the values of $t_1$ and $t^\prime$ increases beyond $2t$, the spectral gap remains trivial ($C = 0$) for all values of $\Delta$.

	The phase diagrams corresponding to the conduction bands are presented in Figs. \ref{fig:pd_t1_1.0_t2_100_ABC_ub}-\ref{fig:pd_t1_1.95_t2_0.1_ABC_ub}. It is evident that they are similar to that of the valence band except that the Chern insulating regions move away symmetrically from $\phi =0$ with these having opposite signs for the Chern number, $C$. In Figs. \ref{fig:pd_t1_1.0_t2_0.1_ABC_mb}-\ref{fig:pd_t1_1.95_t2_0.1_ABC_mb}, the phase diagrams corresponding to the flat band are depicted. It is obvious that the topological regions are away from $\phi = \pi/2$ and both $C = +2$ and $C = -2$ are observed in the $\pi>|\phi|> 0$ regime. Further the combined phase diagrams for the conduction, middle and the valence bands totally fill up the region under the sinusoidal curves that correspond to the band deformed two-band Haldane model \cite{mondal2021}. Hence the total Chern number summed over all bands vanishes. Also we get higher Chern numbers, namely, $|C| = 2$ instead of $|C| = 1$, along with each of the Chern lobes no longer being symmetric (and sinusoidal).
	
	The scenario changes for case-II, when we retain $t^\prime = t$ and selectively vary $t_1$ as shown in Fig. \ref{fig:pd_lb_ub_mb}. In this case the Chern insulating phases are denoted by different colours, namely, cyan for $C = 2$ and red for $C = -2$ phases. It can be noticed that the Chern insulating regions are shifted away from $\phi = 0$ corresponding to the valence band (Figs. \ref{fig:pd_t1_1.2_t2_0.1_lb}-\ref{fig:pd_t1_1.55_t2_0.1_lb}), while it gets shifted away from $|\phi| = \pi$ corresponding to the conduction band (Figs. \ref{fig:pd_t1_1.2_t2_0.1_ub}-\ref{fig:pd_t1_1.55_t2_0.1_ub}). Further, the Chern lobes are sinusoidal in shape and \textbf{thus} are distinct from those in case-I. However as earlier, the Chern insulating regions gradually shrink with increase in $t_1$ and it eventually vanishes completely at the gap closing point $t_1 = 1.67t$. For any non-zero value of $\Delta$, the spectral gap is always trivial for $t_1 = 1.67t$. For $t_1>1.67t$, the gap reopens for $\Delta = 0$, however the Chern number vanishes. Thus we observe a topological phase transition at the gap closing point $t_1 = 1.67t$. This is analogous to the semi-Dirac limit for case-II, which now occurs for lower values of $t_1$. Further, the non-trivial regions corresponding to the middle band have values both $+2$ and $-2$ for $\pi>|\phi|> 0$ (Figs. \ref{fig:pd_t1_1.2_t2_0.1_mb}-\ref{fig:pd_t1_1.55_t2_0.1_mb}). The combined phase diagrams corresponding to the conduction, middle and the valence bands
	again account for the total Chern number to be vanishing and is similar to the previous case.
	
	There is a subtle issue regarding the non-trivial phases of the system which needs to be mentioned in some details (we have made a brief mention of it in Sec. \ref{sec:bandstructure}). For the well known Haldane model, the topological phase of the system does not depend upon the value of the NNN hopping amplitude $t_2$, that is, it only requires an infinitesimal $t_2$ to break the time-reversal invariance, which yields a non-zero Chern number. However, in case of a band deformed dice lattice the non-trivial topology depends on the value of $t_2$. For example, let us fix $t_1$, say at $t_1 = 1.4t$ and vary $t_2$. At $t_2 = 0.05t$, we observe vanishing of the Chern number regardless of the values of $\phi$ and $\Delta$. However, when $t_2 = 0.1t$ we obtain the $C = |2|$ phase for certain range of values for $\phi$ and $\Delta$. This implies that at a particular value of $t_1$, the Chern insulating regions in the $\Delta$-$\phi$ phase diagram gradually increases with increase in the value of $t_2$. However, such variation with $t_2$ occurs for all values of $t_1$, such that $t_1>t$ (except for $t_1 = t$). In Fig. \ref{fig:C_vs_t1} we have plotted the Chern number as a function of the hopping strength $t_1$ for various representative values of $t_2$ indicated in the figure. As can be seen, with increase in the value of $t_2$, the Chern number vanishes at higher values of $t_1$. For example, when $t_2 = 0.02t$, the phase transition occurs at $t_1 \simeq 1.16t $. Whereas, for $t_2 = 0.1t$, the transition occurs at $t_1 \simeq 1.67t$. We quote a few other pair of values of $t_1$ and $t_2$ for which the Chern number vanishes, such as, $(t_1, t_2) \simeq (1.37t, 0.05t)$, $(1.56t, 0.08t)$, $(1.77t, 0.12t)$ and $ (1.84t, 0.14t)$ and so on.
	
\subsection{Edge states}\label{sec:edge_states}
	\begin{figure}[h]
		\captionsetup[subfigure]{labelformat=nocaption}
		\centering
		\includegraphics[width=\linewidth]{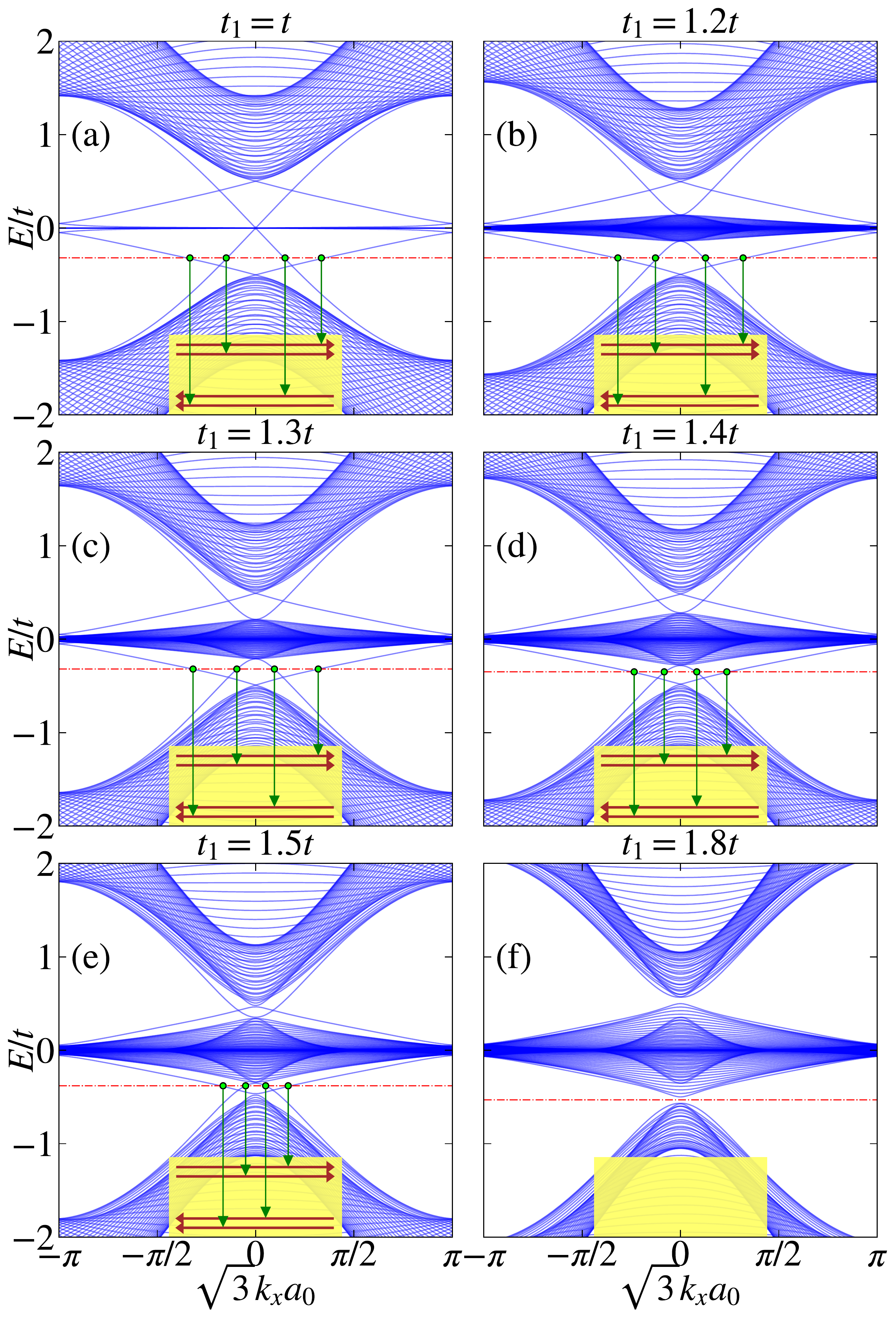}
		\begin{subfigure}[b]{0\textwidth}
			\subcaption{}\label{fig:ES_t1_1_t2_0.1_m_0_phi_0.5}
		\end{subfigure}
		\begin{subfigure}[b]{0\textwidth}
			\subcaption{}\label{fig:ES_t1_1.2_t2_0.1_m_0_phi_0.5}
		\end{subfigure}
		\begin{subfigure}[b]{0\textwidth}
			\subcaption{}\label{fig:ES_t1_1.3_t2_0.1_m_0_phi_0.5}
		\end{subfigure}
		\begin{subfigure}[b]{0.\textwidth}
			\subcaption{}\label{fig:ES_t1_1.4_t2_0.1_m_0_phi_0.5}
		\end{subfigure}
		\begin{subfigure}[b]{0.\textwidth}
			\subcaption{}\label{fig:ES_t1_1.5_t2_0.1_m_0_phi_0.5}
		\end{subfigure}
		\begin{subfigure}[b]{0.\textwidth}
			\subcaption{}\label{fig:ES_t1_1.8_t2_0.1_m_0_phi_0.5}
		\end{subfigure}
		\caption{\raggedright The band structure of the semi-infinite ribbon is shown for (a) $t_1 = t$, (b) $t_1 = 1.2t$, (c) $t_1 = 1.3t$, (d) $t_1 = 1.4t$, (e) $t_1 = 1.5t$, and (f) $t_1 = 1.8t$.  The Fermi level is shown via the red dashed line which intersects the edge modes at four distinct points (shown by the green dots). Corresponding to those intersecting points, the edge currents are shown by the red arrows in the yellow panel at the bottom of each figure. The values of $t^\prime$, $t_2$, $\phi$ and $\Delta$ are taken as $t$, $0.1t$, $\pi/2$ and zero respectively.}
		\label{fig:ES}
	\end{figure}
	
	
	\begin{figure}[h]
		\centering
		\includegraphics[width=0.8\linewidth]{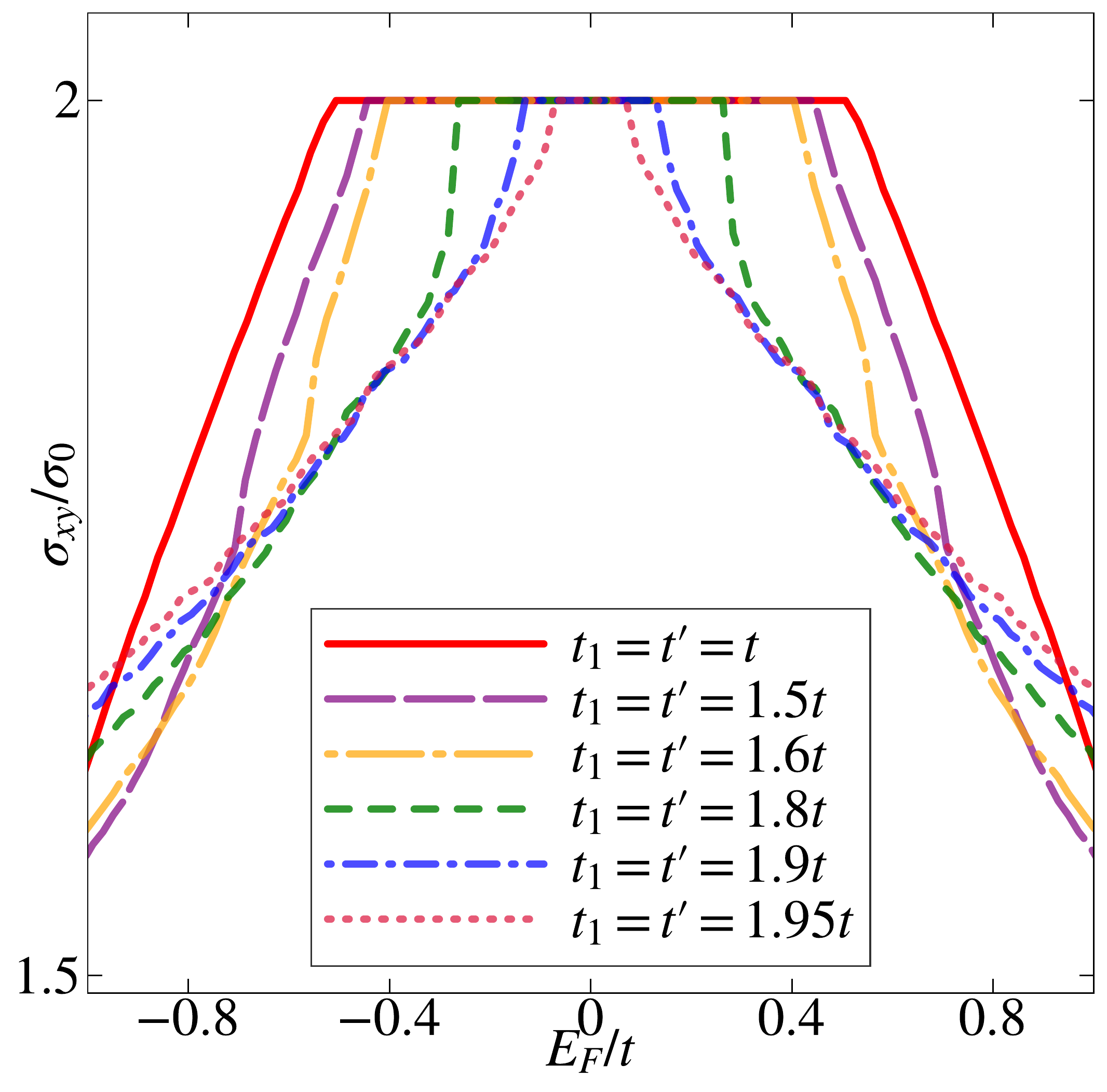}
		\caption{\raggedright The Hall conductivity is depicted for various values of $t_1$ and $t^\prime$ shown in the inset. The other parameters are taken as, $t_2 = 0.1t$, $\phi = \pi/2$ and $\Delta = 0$.}
		\label{fig:hall_cond1}
	\end{figure}
	\begin{figure}[h]
		\centering
		\includegraphics[width=0.8\linewidth]{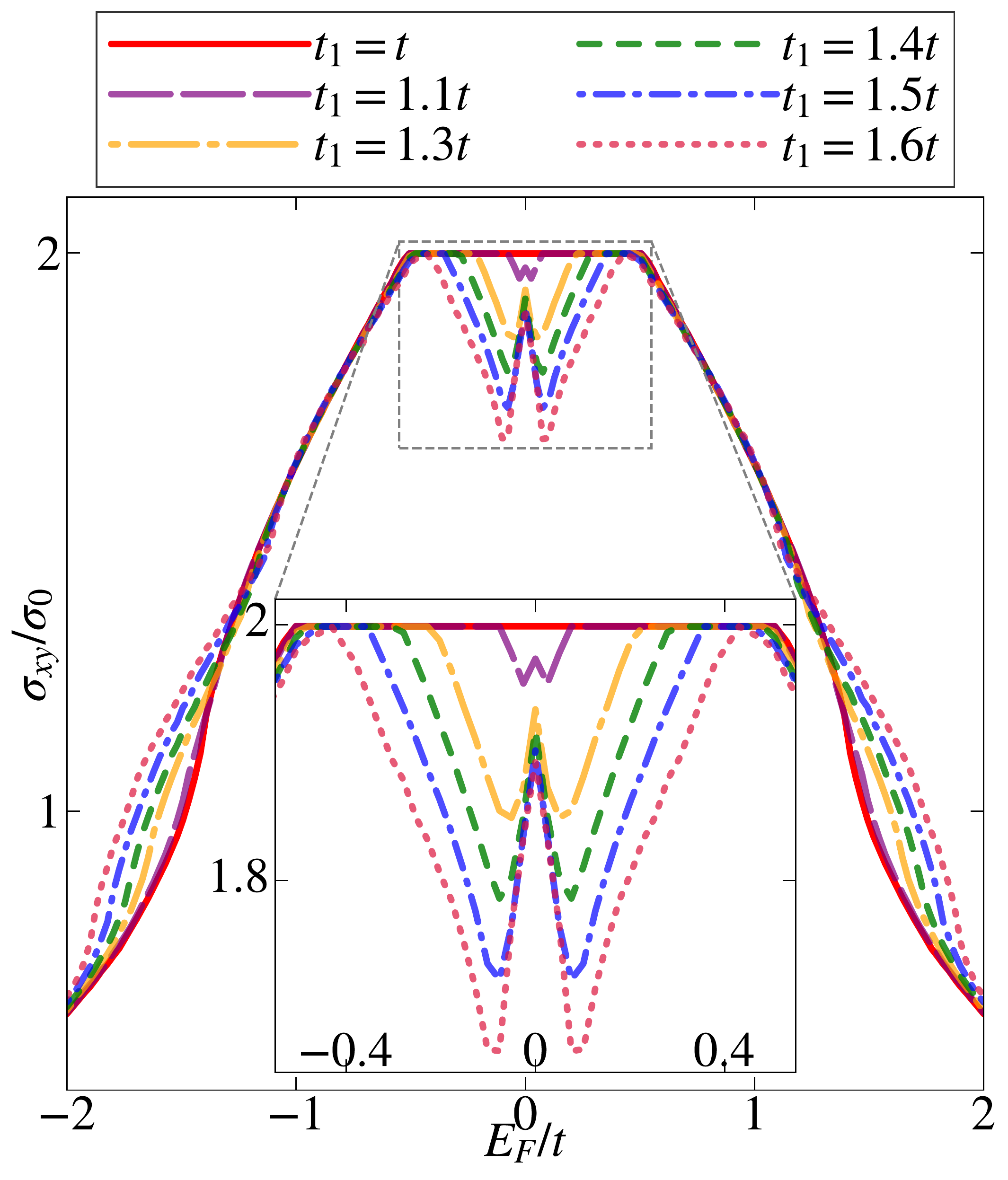}
		\caption{\raggedright The Hall conductivity for a fixed $t^\prime$ is depicted for various values of $t_1$ shown in the panel above. In the inset, a broader view of the regions near zero Fermi energy are shown. The dips in the Hall conductivity are clearly visible. The values of $t^\prime$, $t_2$, $\phi$ and $\Delta$ are fixed at $t$, $0.1t$, $\pi/2$ and zero respectively.}
		\label{fig:hall_cond2}
	\end{figure}
	\begin{figure}[h]
		\captionsetup[subfigure]{labelformat=nocaption}
		\centering
		\includegraphics[width=\linewidth]{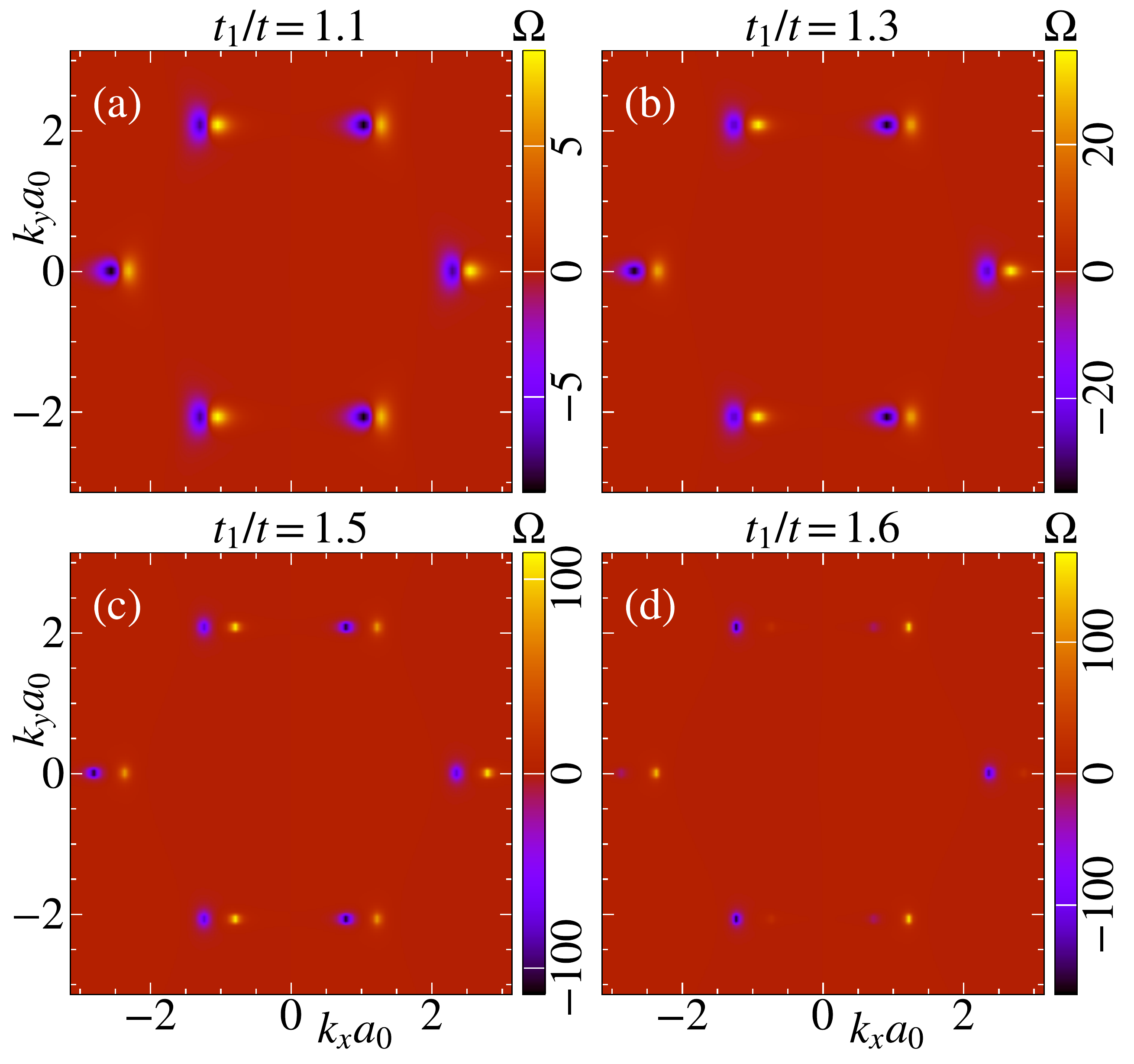}
		\begin{subfigure}[b]{0\textwidth}
			\subcaption{}\label{fig:BC_t1_1.1_t2_0.1_m_0_phi_0.5}
		\end{subfigure}
		\begin{subfigure}[b]{0\textwidth}
			\subcaption{}\label{fig:BC_t1_1.3_t2_0.1_m_0_phi_0.5}
		\end{subfigure}
		\begin{subfigure}[b]{0\textwidth}
			\subcaption{}\label{fig:BC_t1_1.5_t2_0.1_m_0_phi_0.5}
		\end{subfigure}
		\begin{subfigure}[b]{0.\textwidth}
			\subcaption{}\label{fig:BC_t1_1.6_t2_0.1_m_0_phi_0.5}
		\end{subfigure}
		\caption{\raggedright The Berry curvature corresponding to the middle band is presented for (a) $t_1 = 1.1t$, (b) $t_1 = 1.3t$, (c) $t_1 = 1.5t$ and (d) $t_1 = 1.6t$. The values of $t^\prime$, $t_2$, $\phi$ and $\Delta$ are again taken as $t$, $0.1t$, $\pi/2$ and zero respectively.}
		\label{fig:BC}
	\end{figure}

	In order to visualize the nature of the band gap, that is, whether it is topological or trivial, we look for the crossings of the edge modes with the Fermi energy. To obtain such edge modes in our calculation, we consider a semi-finite ribbon of the system which breaks the periodicity along one direction, while the translational symmetry remains intact along the perpendicular direction. We take the ribbon \cite{nakada1996, sticlet2012} to be finite along the $y$-direction and infinite along the $x$-direction with armchair edges. Further, the sites along the $y$-direction are labelled as A$_1$, B$_1$, C$_1$, A$_2$, B$_2$, C$_2$ .... A$_N$, B$_N$, C$_N$ etc. Now we Fourier transform the operators along the $x$-direction ($k_x$ being a good quantum number). This yields following sets of three coupled eigenvalue equations for the wave function amplitudes. 
	\begin{equation}\label{eq:edge1}
		\begin{aligned}
			E_{k} a_{k, n} =&\left[ t \left\{ b_{k, n+1} + b_{k, n-1} \right\} + t_1 b_{k, n} e^{-i\eta_1 k} \right] + \Delta a_{k, n}\\ & t_2\left[e^{+i\phi} \left\{ a_{n -2} + a_{n +1}\left( 1 + e^{i (-1)^{n + 1}k} \right) \right\} \right. \\ & \left. e^{-i\phi} \left\{ a_{n+2} + a_{n-1}\left( 1 + e^{i (-1)^{n + 1}k}\right) \right\}\right]
		\end{aligned}
	\end{equation}
	\begin{equation}\label{eq:edge2}
		\begin{aligned}
			E_{k} b_{k, n} =&\left[ t \left\{ a_{k, n+1} + a_{k, n-1} \right\} + t_1 a_{k, n} e^{+i\eta_1 k} \right] \\& + \left[ tc_{k, n+1} + tc_{k, n-1} + t^\prime c_{k, n} e^{-i\eta_2 k} \right]
		\end{aligned}
	\end{equation}
	\begin{equation}\label{eq:edge3}
		\begin{aligned}
			E_{k} c_{k, n} =& \left[ tb_{k, n+1} + tb_{k, n-1} + t^\prime b_{k, n} e^{+i\eta_2 k} \right] -\Delta c_{k, n}\\
			& t_2\left[e^{+i\phi} \left\{ c_{n+2} + c_{n-1}\left( 1 + e^{i (-1)^{n + 1}k} \right) \right\} \right. \\ & \left. e^{-i\phi} \left\{ c_{n+1} + c_{n-2}\left( 1 + e^{i (-1)^{n + 1}k}\right) \right\}\right] 
		\end{aligned}
	\end{equation}
	where $a_{k, n}$, $b_{k, n}$ and $c_{k, n}$ are the coefficients of the wave function corresponding to the sublattices A, B and C respectively and $n$ is the site index which takes integer values between $[1:N]$, with $N$ being the total number of unit cells along the $y$-direction. In Eqs. \ref{eq:edge1}, \ref{eq:edge2} and \ref{eq:edge3}, $\eta_1$ and $\eta_2$ are written as $\eta_1=\left\{ 1 + (-1)^{n} \right\}/2$ and $\eta_2=\left\{ 1 + (-1)^{n +1 } \right\}/2$ respectively. Further, the momentum $k$ is the scaled $k_x$ variable and is defined by $k = \sqrt{3}a_0 k_x$, such that it is rendered dimensionless. The width of the ribbon is related $N$ via $D(N) = \frac{3a_0}{2}(N - 1)$. In our calculation we have taken $N = 80$, which gives the width $D(N) = 237a_0/2$. 
	
	By solving Eqs. \ref{eq:edge1}, \ref{eq:edge2} and \ref{eq:edge3}, we have computed the band structure of the semi-infinite ribbon (armchair) corresponding to case-II, that is, for $t^\prime = t$, $t_2 = 0.1t$, $\phi = \pi/2$ and $\Delta = 0$ which is presented in Fig. \ref{fig:ES}. As can be seen, a pair of edge modes from the valence bands crosses over to the middle bands (they are bunched together around the zero energy) and another pair crosses over in the opposite direction. A similar scenario happens corresponding to the modes connecting the conduction bands and the middle bands. Because of these edge modes, the Hall conductivity remains finite, provided the Fermi energy lies in the bulk gap. The Fermi energy in each figure is shown via the red dashed line, which intersects the edge modes at four points (shown by green dots). The edge currents corresponding to those green dots, are shown by the red arrows in the yellow panels in Figs. \ref{fig:ES_t1_1_t2_0.1_m_0_phi_0.5} - \ref{fig:ES_t1_1.5_t2_0.1_m_0_phi_0.5}. The yellow panels represent a part of the semi-infinite ribbon. Therefore, there are a pair of edge currents along either edges of the ribbon. Since the velocity of the electron is proportional to the slope of the bands, that is, $\partial E/\partial k$, the edge currents along a particular edge move in the same direction, however, along the other edge, each pair moves in the opposite direction. These modes are chiral (and not helical).
	
	Owing to the presence of a pair of edge modes, there will be a quantized Hall conductivity occurring at a value $2e^2/h$, with the factor `2' in front denoting the number of edge modes \cite{hatsugai1988}. This result is consistent with the corresponding Chern numbers of the system as presented in Fig. \ref{fig:pd_lb_ub_mb}. 
	For example, the non-zero edge currents are observed for $t_1 < 1.67t$ (see Fig. \ref{fig:ES_t1_1_t2_0.1_m_0_phi_0.5} - \ref{fig:ES_t1_1.5_t2_0.1_m_0_phi_0.5}) and in this range the Chern number is found to be $|C| = 2$. For $t_1>1.67t$, say at $t_1 = 1.8t$, we observe a vanishing Chern number and as well as the edge currents disappear (see Fig. \ref{fig:ES_t1_1.8_t2_0.1_m_0_phi_0.5}). The behaviour of the edge states corresponding to case-I, that is, when $t_1$ and $t^\prime$ are varied simulatneously, are not shown since they are similar to what we observe in Fig. \ref{fig:ES}, with the critical value of $t_1$ being the only difference, that is, the edge modes exist till both $t_1$ and $t^\prime$ remain just below $2t$ (as opposed to $1.67t$).

	\subsection{Hall conductivity}\label{sec:hall_conductivity}

	\begin{figure}[h]
		\captionsetup[subfigure]{labelformat=nocaption}
		\centering
		\includegraphics[width=\linewidth]{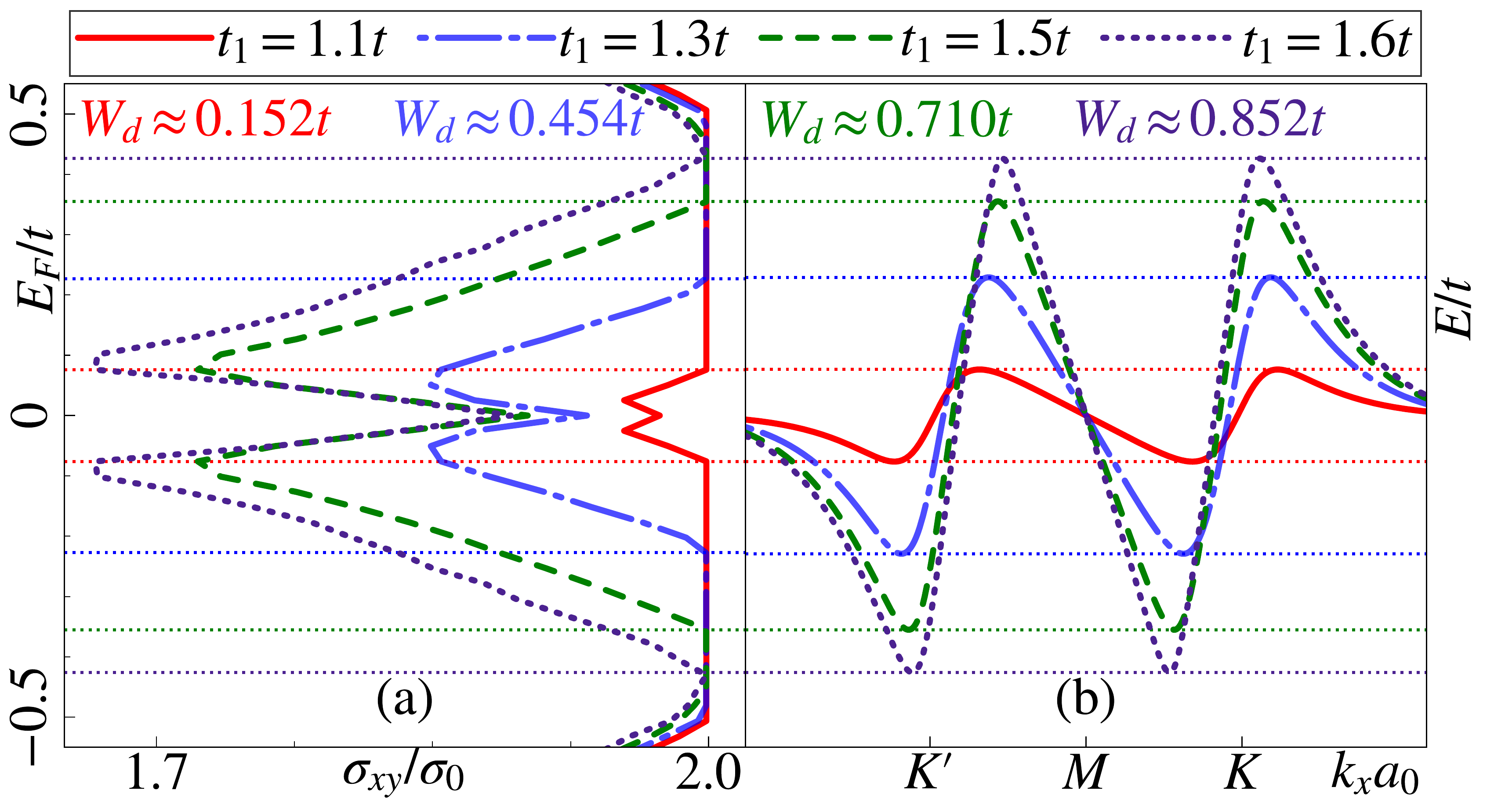}
		\begin{subfigure}[b]{0\textwidth}
			\subcaption{}\label{fig:Hall_Wd}
		\end{subfigure}
		\begin{subfigure}[b]{0\textwidth}
			\subcaption{}\label{fig:Hall_band_Wd}
		\end{subfigure}
		\caption{\raggedright A comparison between the dips in the plateau width ($W_d$) and the bandwidths of the middle band for different values of $t_1$ are shown.  The Hall conductivities are plotted in (a) along the $x$-direction, while the middle bands are depicted in (b). Different colours signify different values of $t_1$. $W_d$ for various values of $t_1$ are mentioned above the figures.}
		\label{fig:Hall_width}
	\end{figure}
	In this section we discuss the Hall conductivity of the system. In order to calculate it, first the Berry curvature needs to be obtained first using Eq. \ref{eq:berry_curv} and then the following formula can be used \cite{hall1,hall2}.
	\begin{equation}\label{eq:Hall_cond}
		\sigma_{xy} = \frac{\sigma_0}{2\pi} \sum_{\lambda} \int \frac{\mathrm{d}k_x \mathrm{d}k_y}{(2\pi)^2} f\left(E^\lambda_{k_x, k_y} \right) \Omega(k_x, k_y)
	\end{equation}
	where $E^\lambda(k_x, k_y)$ is the energy dispersion with $\lambda$ being $+1$, $0$ and $-1$ corresponds the conduction, middle and valence bands respectively. $f(E) = \left[ 1+e^{(E-E_F)/K_BT} \right]^{-1}$ signifies the Fermi-Dirac distribution function with $E_F$ and $T$ being the Fermi energy and absolute temperature respectively. The constant term $\sigma_0 = e^2/h$ sets the scale for $\sigma_{xy}$. Now, the Hall conductivity can be obtained numerically at the zero temperature as a function of the Fermi energy $E_F$ as shown in Figs. \ref{fig:hall_cond1} and \ref{fig:hall_cond2}. 
	
	First we discuss the spectrum for the variation of both $t_1$ and $t^\prime$ (case-I) which is presented in Fig. \ref{fig:hall_cond1}. It can be noticed that a plateau quantized at $2e^2/h$ exists as long as $E_F$ remains in the bulk gap of the dispersion spectrum. The value of $\sigma_{xy}$ starts to decrease when $E_F$ leaves the bulk gap and goes inside the band (either conduction or the valence bands). This results in diminishing of the plateau width with increase in value of $t_1$ and $t^\prime$, since the gap between the conduction and valence bands shrinks. The Hall conductivity vanishes completely when both $t_1$ and $t^\prime$ become larger than $2t$. Thus, we observe plateau at $|C|e^2/h$ as long as the system remains a non-trivial insulator, which vanishes at the gap closing point at $t_1 = t^\prime = 2t$.
	

	In Fig. \ref{fig:hall_cond2}, we have presented the Hall conductivity as a function of $E_F$ for case-II where only $t_1$ is tuned, while $t^\prime$ is fixed at $t$. In this case, the middle band plays an important role in the behaviour of the Hall conductivity as it acquires a dispersive nature.
	For example, let us fix the value of $t_1$ at $1.1t$, for which the variation of $\sigma_{xy}$ is shown in Fig. \ref{fig:hall_cond2} via a purple colored curve. It shows that there is no longer a smooth plateau in the Hall conductivity, instead, $\sigma_{xy}$ acquires a dip in the vicinity of zero bias along with a spike occurring at $E_F\simeq 0$. The spike gets more and more prominent with increase  in the value of $t_1$.
	Further, the dips widen with the increase of $t_1$ resulting in a diminishing width of the plateau region around the zero Fermi energy. Let us denote the width of the dip as $W_d$. Notably $W_d$ becomes very large for $t_1 = 1.6t$. Finally, the plateau and hence the Hall conductivity vanish completely for $t_1\gtrsim 1.67t$.
	
	The reason of getting a dip in the Hall conductivity can be inferred from the fact that the dispersive middle band has non-zero Berry curvature $\Omega$ (see Fig. \ref{fig:BC}) which would otherwise be zero for the flat band. Let us say, we set $E_F$ in such a way that some of the states corresponding to the middle band lies above $E_F$ and some lie below it. So, when we compute the integral of $\Omega$ over the occupied states corresponding to the middle band, we get a non-zero value. However, when it is included in the contributions from that of the valence band, the result yields lesser values than $2\sigma_0$ ($\sigma_0=e^2/h$) for the Hall conductivity. On the other hand, when $E_F$ lies between the gap of the middle and the conduction bands, the integral of $\Omega$ corresponding to the middle band completely vanishes. This is consistent with the corresponding value of the Chern number, which is zero for the middle band at $\phi = \pi/2$. Therefore, we observe a plateau at $2\sigma_0$ when $E_F$ lies in the band gap.

	In Fig. \ref{fig:Hall_width}, we have compared the width of the dip ($W_d$) in the Hall conductivity with the bandwidth of the middle band. The figure clarifies that $W_d$ scales with the bandwidth of the middle  band. $W_d$ becomes zero for $t_1 = t$. It increases with increase of $t_1$, and $W_d$ is maximum below $t_1 = 1.67t$, where the middle band touches both the valence and conduction bands. Beyond $t_1 = 1.67t$, the Hall conductivity vanishes completely.

	\section{Conclusion}\label{sec:conclusion}
	
	We have presented two schemes through which hopping anisotropies are introduced that induce a band deformation of a Haldane model on dice lattice. In case-I, both $t_1$ (A-B hopping) and $t^\prime$ (B-C hopping) are varied where the band extrema from the $\mathbf{K}$ and $\mathbf{K}^\prime$ points are shifted. Along with that the band gap diminishes which finally vanishes in the semi-Dirac limit, namely, at $t_1 =t^\prime= 2t$. In contrast, case-II refers to a situation where only $t_1$ is varied, and the spectrum shows dispersive flat bands. Unlike case-I, the band extremum located at one of the Dirac points is displaced slightly and the spectral gap closes at $t_1 = 1.67t$ in case-II. In both the cases we observe the Chern numbers $\pm 2$ till their respective gap closing transition occurs. However, the shape of the topological lobes in the phase diagrams are distinct in these two cases. Further, a pair of chiral edge modes at each edge are obtained in a nanoribbon geometry which confirms the values of the Chern numbers obtained in these two cases. We have also calculated the anomalous Hall conductivity, which shows plateaus quantized at $2e^2/h$ as long as $t_1$ and $t^\prime$ remain less than $2t$ corresponding to case-I. While in case-II, because of the finite Berry curvature of the middle band, a dip in the plateau appears for $t_1\neq t$ close to zero Fermi energy. Such dip widens (or the plateau width decreases) with increasing of $t_1$. The Hall conductivity vanishes for $t_1\geq 1.67t$ which is in agreement of the vanishing Chern numbers at such values of $t_1$. Thus, our models of band deformed dice lattice present a topological phase transition from a Chern insulating phase to a trivial insulator across a gap closing point for the hopping amplitude values given by $(t_1 , t^\prime) = (2t, 2t)$ in case-I and $(t_1 , t^\prime) = (1.67t, t)$ in case-II.

\end{document}